\documentclass[aps,prd,10pt,twocolumn,
unsortedaddress,
amsfonts,amssymb,amsmath,byrevtex,showpacs,nofootinbib]{revtex4-1}

\usepackage{latexsym}
\usepackage{graphicx}
\usepackage{geometry}
\usepackage{epstopdf}
\usepackage{color}

\usepackage{xcolor}
\usepackage{pdfsync}
\usepackage{fancyhdr}
\usepackage[applemac]{inputenc}
\usepackage[breaklinks,colorlinks,backref]{hyperref}
\hypersetup{
    colorlinks, 
    linktoc=all, 
    linkcolor=red,  
  citecolor=hyptxt,
  urlcolor=blue}
  \hypersetup{
  citebordercolor=,
  filebordercolor=green,
  linkbordercolor=blue
}
\definecolor{hyptxt}{rgb}{0.7, 0.4, 0.9}
\usepackage{bibtopic}

\newcommand{\ud}{\mathrm{d}}
\newcommand{\be}{\begin{equation}}
\newcommand{\ee}{\end{equation}}

\newcommand{\N}{\mathbb N}

\newcommand{\R}{\mathbb R}

\newcommand{\ket}[1]{|\kern.3ex#1\kern.3ex\rangle}
\newcommand{\bra}[1]{\langle\kern.3ex #1 \kern.3ex|}
\newcommand{\scalar}[2]{\langle\kern.3ex #1 \kern.3ex|\kern.3ex#2\kern.3ex\rangle}
\newcommand{\norm}[1]{\|\kern.3ex#1\kern.3ex \|}

\def\red{\textcolor{red}}

\def\lg{\langle }
\def\rg{\rangle }

\def\ud{\mathrm{d}}
\def\mfn{\mathfrak{n}}

\begin{document}

\title{Singularity avoidance in a quantum model of the Mixmaster universe}

\author{Herv\'{e} Bergeron}
\email{herve.bergeron@u-psud.fr} \affiliation{ISMO, CNRS, Universit\'{e}~Paris-Sud, Universit\'{e} Paris-Saclay, 
91405 Orsay, France}

\author{Ewa Czuchry}
\email{eczuchry@fuw.edu.pl} \affiliation{National Centre for Nuclear Research, Ho{\.z}a 69,
00-681 Warszawa, Poland}

\author{Jean-Pierre Gazeau}
\email{gazeau@apc.univ-paris7.fr}
\affiliation{APC, UMR 7164 CNRS, Universit\'{e} Paris Diderot, Sorbonne Paris Cit\'e, 75205 Paris, France}
\affiliation{Centro Brasileiro de Pesquisas Fisicas
22290-180 - Rio de Janeiro, RJ, Brazil }

\author{Przemys{\l}aw Ma{\l}kiewicz}
\email{pmalkiew@fuw.edu.pl}
\affiliation{National Centre for Nuclear Research,  Ho{\.z}a 69,
00-681
Warszawa, Poland}
\affiliation{APC, Universit\'e Paris Diderot, Sorbonne Paris Cit\'e, 75205 Paris
Cedex 13, France}

\author{W{\l}odzimierz Piechocki}
\email{wlodzimierz.piechocki@ncbj.gov.pl} \affiliation{National Centre for Nuclear Research, Ho{\.z}a
69, 00-681 Warszawa, Poland}

\date{\today}


\begin{abstract}
We present a quantum model of the vacuum Bianchi-IX dynamics. It is based on four main elements. First, we use
a compound quantization procedure: an affine coherent state quantization
for isotropic variables and a Weyl quantization for anisotropic ones. Second, inspired by standard approaches in molecular physics, we make an adiabatic approximation (Born-Oppenheimer-like approximation). Third, we expand the anisotropy potential about its minimum in order to deal with its harmonic approximation. Fourth, we develop an analytical treatment on the semi-classical
level. The resolution of the classical singularity occurs due to a
repulsive potential generated by the affine quantization. This
procedure shows that during contraction the quantum energy of
anisotropic degrees of freedom grows much slower than the
classical one. Furthermore, far from the quantum bounce, the classical re-collapse is reproduced.  Our 
treatment is put in the general context of methods of molecular 
physics, which can include both adiabatic and non-adiabatic approximations.
\end{abstract}


\maketitle


\section{Introduction}

The Belinskii, Khalatnikov and Lifshitz (BKL)
scenario \cite{BKL1,BKL2} (see \cite{Gar} for numerical support
for BKL),  addresses the generic solution to the
Einstein equations near the cosmological singularity. The purpose of our paper
is to quantize the dynamics of the vacuum Bianchi-IX model that
underlies the BKL scenario.

The BKL predicts that on approach to a spacelike singularity
the dynamics of gravitational field may be significantly
simplified as time derivatives in Einstein's equations dominate
over  spatial derivatives.  The latter means that the evolution of
the gravitational field in this regime is ultralocal and space
splits into collection of small patches whose dynamics is
approximately given by spatially homogenous spaces, the Bianchi
models (see, e.g. \cite{Bojowald:2003xe}). Approaching the
singularity the spatial curvature grows and the space further
subdivides into homogenous slices. The size of each patch,
modeled by one of the Bianchi spacetimes, corresponds to the
magnitude of the spatial derivatives in the Einstein equations. As
homogeneity of  spatial fragments holds only at some level of
approximation, dynamical evolution of the newly formed patches
starts off with slightly different initial conditions.  This almost
negligible difference grows rapidly in subsequent evolution as
geometries of new patches evolve almost independently of each
other approaching the state of the so-called asymptotic silence
\cite{Andersson:2004wp}. The chaotic subdivisions result in the
growth of fragmentation of spacetime suggesting that it may be
possessing fractal structure close to the singularity
\cite{Cornish:1996yg,Cornish:1996hx}.

Among the possible homogeneous models, the Bianchi-IX model has
sufficient generality to describe the evolution of a small patch
of space towards the singularity. The dynamics of the vacuum
Bianchi-IX model (i.e., the mixmaster universe) is nonintegrable.
However, close enough to the singularity, each solution
can be qualitatively understood as a sequence of Kasner epochs,
which correspond to the Kasner universe. The transitions between
the epochs are described by the vacuum Bianchi-II type evolution
\cite{Bergeron:2014kea}. The universe undergoes an infinite number
of chaoticlike transitions and eventually collapses into the
singularity in a finite proper time \cite{BKL1}.

The imposition of quantum rules into the chaotic dynamics of the
Bianchi-IX model has been already studied.  The program initiated
by Misner \cite{cwm1,cwm2,cwm3} led to the pessimistic result that
quantum mechanics does not remove the singularity of the Bianchi-IX
model.  Nevertheless, the exploration of solutions to the
corresponding Wheeler-DeWitt equation continues \cite{Marolf}, \cite{Moncrief}, \cite{Bae}, \cite{Koehn}.
Recently, some effort has been made towards quantization of the Bianchi-type
models within the loop quantum cosmology. The authors make use of the Dirac quantization
method and combine it with the introduction of holonomies in place
of the curvature of connection. The results obtained for the
Bianchi-IX model at semiclassical level by Bojowald
\cite{Bojowald:2003xe,Bojowald:2004ra} suggest that the chaotic
behavior stops once quantum effects become important. Another
formulation taking into account holonomies has been proposed  in
\cite{WilsonEwing:2010rh}, but it has not been applied to the
examination of the dynamics. Still another proposal was given in
\cite{Lecian:2013rea}, giving support to \cite{Bojowald:2004ra}.
 An effective dynamics considered recently \cite{Corichi:2015ala}
suggests a resolution of the cosmic singularity problem as well, but
within an approach quite different from ours.
In the above formulations, the search for solutions is quite
challenging leaving the near big bang dynamics largely unexplored.

In  this paper we formulate and make a quantum study of
Bianchi-IX model by combining canonical and affine coherent state
(ACS) quantizations with a semiclassical approach. The cosmological
system consists of isotropic variables (expansion and volume) and
anisotropic ones (distortion and shear). They are treated in a
separate manner. The canonical pair expansion volume is a
half plane. Since the  symmetry of the latter is the affine group
``$AX+B$'' and not the Abelian $\R^2$, they are consistently quantized by resorting to one of
the two unitary irreducible representations (UIRs) of the affine group. Within this approach, we have found in
\cite{Bergeron:2013ika} that for the Friedmann-Robertson-Walker
(FRW) models the cosmological expansion squared, which plays the
role of kinetic energy of the universe, is always accompanied at the
quantum
level by an extra term inversely proportional to the volume squared.
As the Universe approaches the singularity, this term grows in
dynamical significance, efficiently counterbalances the attraction
of any matter and eventually halts the cosmological contraction.
Afterwards, the Universe rebounds and re-expands. In the present
work (see also \cite{bergeronshort}) we confirm that the same
mechanism prevents the collapse of Bianchi-IX universe, suggesting
its universality. Making further use of the 
ACS we construct a semiclassical description of the isotropic
part of the metric, with semiclassical observables replacing the
classical ones in the phase space. In particular, the
semiclassical Hamiltonian possesses the correction term, which
regularizes  the singularity.

Inspired by standard approaches in molecular physics, we make an
assumption about the quantum evolution of the anisotropic
variables based on the adiabatic approximation. In molecules, the
motion of heavy nuclei is so slow in comparison with rapidly
moving light electrons that it is legitimate to approximate the
dynamics with electronic configurations being instantaneously and
continuously adjusted to the position of nuclei. Analogously, we
consider in our model, in its harmonic approximation, the anisotropic oscillations rapid in
comparison with the contraction rate of the Universe. Within this
approach, the oscillations of the classical scenario are
suppressed and the development of chaos is blocked. Moreover, we
find that while the classical energy of the oscillations behaves
in terms of the scale factor $a$ more or less as $\propto a^{-6}$,
the respective quantum energy behaves as $\propto a^{-4}$, i.e. it
contributes on a much softer level. Therefore, fluids with
pressure equal or higher than that of radiation, which are likely
to be present in the early universe, will have their grasp on the
cosmological collapse.

The paper is organized as follows: Section \ref{bianchiA} concerns
the definition of the classical model, some description of its dynamics
and the choice of phase space variables convenient for our
quantization. Section  \ref{quantiz}  is devoted to the
quantization of the Hamiltonian constraint and its subsequent
semi-classical approximation. Subsection \ref{appsclag} explains the semiclassical Lagrangian approach. The resulting  semiclassical
dynamics is developed in Section \ref{hamcons}. In Section
\ref{consistency}, we go beyond the adiabatic approximation to
confirm the validity of our method. We discuss our results and
conclude  in Sec \ref{conclusion}. In Appendix \ref{affquant} we
give an introduction to the affine coherent states quantization together with its semi-classical aspects. Derivations of the quantum version of the
anisotropic Hamiltonian both in harmonic and triangular box
approximations are given in Appendix \ref{appanqh}.

\section{Bianchi-type models}
\label{bianchiA}

We consider a spacetime admitting a foliation $\mathcal{M}\mapsto
\Sigma\times\mathbb{R}$, where $\Sigma$ is spacelike. Furthermore,
we assume $\Sigma$ to be identified with a simply transitive
three-parameter group of motions. Such models are called Bianchi
type models. The left-invariant vector fields are associated with
the Killing vectors and the right-invariant ones with the basis
vectors with respect to which the metric components on $\Sigma$
take constant values in space. We assume the following line element
\begin{equation}\label{eq}
\ud s^2= - \mathcal{N}^2(t)\ud t^2+\sum_i
q_i(t){\omega^i}\otimes{\omega^i}\, ,
\end{equation}
where the $\omega^i$'s are right-invariant dual vectors. They satisfy
\begin{equation}\label{cartan}
\ud\omega^i=\frac{1}{2}C^i_{~jk}\omega^j\wedge\omega^k\, ,
\end{equation}
where $C^i_{~jk}$ are structure constants. We consider the so-called
Class A models with $C^i_{~ik}=0$ (summation is implied). The further
simplification is gained for the diagonal ones,
$C^i_{~jk}=\mfn^{(i)}\epsilon_{(i)jk}$, where $\epsilon_{ijk}$ is
a totally skew-symmetric symbol. For such models, the computation of
the Ricci curvature is straightforward:
\begin{align}
\nonumber R=&\frac{\mfn^1\mfn^2}{q_3}+\frac{\mfn^1\mfn^3}{q_2}+\frac{\mfn^2\mfn^3}{q_1}
-\frac{(\mfn^1)^2}{2}\frac{q_1}{q_2q_3}\\
&-\frac{(\mfn^2)^2}{2}\frac{q_2}{q_1q_3}
-\frac{(\mfn^3)^2}{2}\frac{q_3}{q_1q_2}
\end{align}
The vector $(\mfn_1,\mfn_2,\mfn_3) \in \mathbb{R}^3$ specifies the Bianchi-type model  (I, II, $\textrm{VI}_0$, $\textrm{VII}_0$, VIII, IX).
Conventionally, the $\mfn_i$'s are chosen as $\mfn_i \in
\{0,\pm1\}$  \cite{EMac}. Special cases are $\mfn_i=0$ (type I) and
$\mfn_i> 0$ (type IX). From now on, we fix $\mfn_i=\mfn$ for the
Bianchi-IX case examined in this work.
We assume that the topology of the spatial leaf is $S^3$ and we find its
coordinate volume as:
\begin{equation}
\mathcal{V}_0=\int_{S^3}\omega^1\wedge\omega^2\wedge\omega^3=\frac{16\pi^2}{\mfn^3}
\end{equation}
Two convenient choices are either $\mfn=1$ or $\mfn=\sqrt[3]{16\pi^2}$.
We  make use of the latter option.

\subsection{Canonical formulation}

Following the work of Arnowitt, Deser and Misner  \cite{adm}, the convenient formulation
of Bianchi models was derived by Misner
\cite{cwm1,cwm2,cwm3}.  With Misner's variables the
dynamics assumes  a convenient form: motion of a particle in
three-dimensional Minkowskian space-time and in a
space-and-time-dependent confining potential. The spatial
coordinates describe the anisotropic distortion of the shape of
Universe and the time coordinate describes the size of Universe.
The particle motion is ruled by a potential arising from the Ricci
curvature of spatial leaf. Let us recall that the Hamiltonian
constraint reads
\begin{align}
\label{Hamilphys00}  {\sf H}=&\frac{\mathcal{N} \mathcal{V}_0 e^{-3\beta_0}}{48\kappa} \times\\
 \nonumber &\left(-\tilde{p}_0^2+\tilde{p}_+^2+
\tilde{p}_-^2
-24e^{6\beta_0}R(\beta_0,\beta_{\pm})\right),
\end{align}
where the Misner configuration variables are related to the metric
components as follows:
\begin{equation}
\left(\begin{array}{c}\ln q_1 \\ \ln q_2 \\ \ln q_3\end{array}\right)=
\left(\begin{array}{ccc}2 & 2 & 2\sqrt{3} \\ 2 & 2 & -2\sqrt{3} \\
2 & -4 & 0\end{array}\right)\left(\begin{array}{c}\beta_0 \\\beta_+
\\ \beta_-\end{array}\right)
\end{equation}
and where $\tilde{p}_0,\tilde{p}_+$ and $\tilde{p}_-$ are the respective
momenta,  defined from the Poisson brackets in Eq. \ref{eq:poissonbracket}.
$\mathcal{V}_0$ is
the coordinate volume and $\kappa=8\pi G c^{-4}$.  The momenta $\tilde{p}_i$
carry the dimension $L^{-1}$ while the positions $\beta_i$ are dimensionless.
Because we reduce a field theory to a mechanical system all the
canonical variables are in fact averaged over the sphere and the Poisson
brackets read as:
\begin{equation}
\label{eq:poissonbracket}
\{\beta_0 , \tilde{p}_0\}=\{\beta_{\pm},\tilde{p}_{\pm}\}=\frac{2\kappa c}{\mathcal{V}_0}
\end{equation}
In order to work within the standard quantum mechanical framework,
we introduce variables that are conjugated in the usual sense,
i.e. $\{\beta_i, p_j\} = \delta_{ij}$, and so define the new
momenta $p_i = (2 \kappa c)^{-1} \mathcal{V}_0 \tilde{p}_i$. The
Hamiltonian of Eq. \eqref{Hamilphys00} now reads:
\begin{align}
\label{Hamilphys0}  &{\sf H}= - \frac{\mathcal{N} \mathcal{V}_0 e^{-3\beta_0}}{48\kappa} \times \\
 \nonumber & \left(\left( \frac{2 \kappa c}{\mathcal{V}_0} \right)^2 \left(p_0^2-p_+^2-p_-^2\right)+
24e^{6\beta_0}R(\beta_0,\beta_{\pm})\right),
\end{align}

In what follows we put $\mathcal{V}_0=2\kappa=c=1$ (in
\cite{bergeronshort} we chose $\kappa=1$). The physical constants
$\mathcal{V}_0$, $\kappa$ and $c$ will be eventually restored  in
the presentation of the final results. Note that the averaged
(isotropic) scale factor ``$a$'' is defined as $a=
e^{\beta_0}=(q_1q_2q_3)^{1/6}$. Observe also that the usual
diffeomorphism constraints vanish identically.

In the case of Bianchi-IX geometry we find
\begin{align}\label{b9curvature}
R(\beta_0,\beta_{\pm})&=-\frac{\mfn^2}{2}e^{- 2\beta_0+
4\beta_+} \times \\
\nonumber & \left(\left[2\cosh(2\sqrt{3}\beta_-)-e^{- 6\beta_+}
\right]^2-4\right)\\
&=-\frac{3}{2}e^{- 2\beta_0} W_{\mfn}(\beta) \,,
\end{align}
with
\begin{align}
\label{b9potential}
W_n(\beta) &= \mfn^2 \frac{e^{4\beta_+}}{3} \times\\
\nonumber & \left(\left[2\cosh(2\sqrt{3}\beta_-)-e^{- 6\beta_+}
\right]^2-4\right) \,.
\end{align}
where the potential $W_{\mfn}$ does not depend on the averaged scale factor $a$.
By putting $p_{\pm}=0=\beta_{\pm}$ we retrieve the
closed Friedmann-Robertson-Walker model, with $W_{\mfn}(0)=- \mfn^2$ giving
rise to the isotropic and positive intrinsic curvature. For $\mfn=0$, we get
the Bianchi I model with $W_{0}=0$ and vanishing intrinsic curvature.

\begin{figure}[!ht]
\includegraphics[scale=0.8]{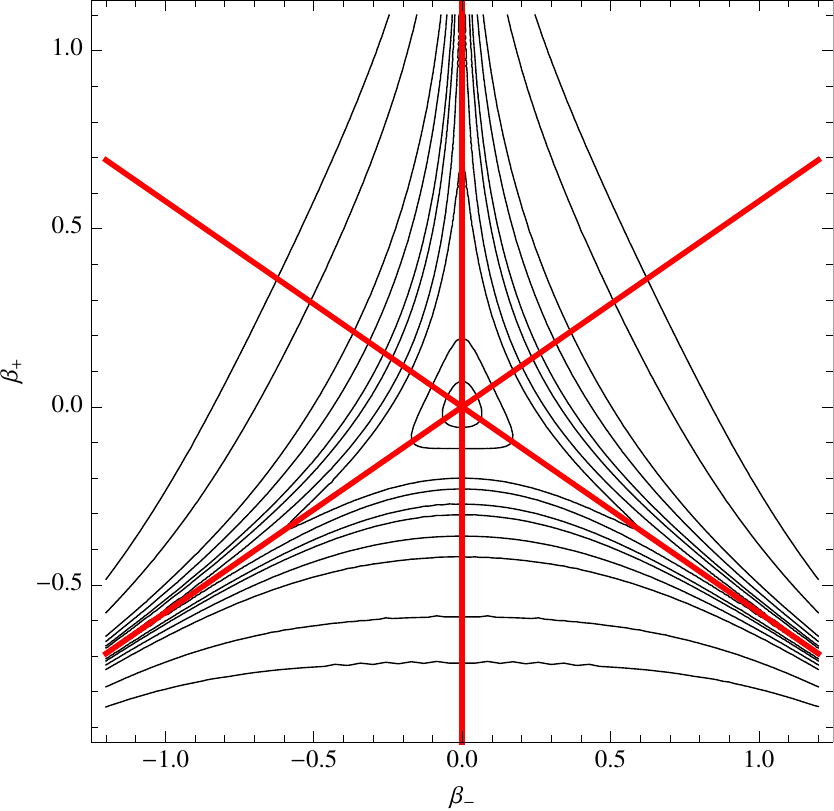}
\caption{Contour plot of $W_{\mfn}(\beta)$ near its minimum. Red
lines: the three ${\sf C}_{3v}$ symmetry axes $\beta_-=0$,
$\beta_+=\beta_-/\sqrt{3}$, $\beta_+=-\beta_-/\sqrt{3}$.}
\label{figure1}
\end{figure}

\begin{figure}[!ht]
\includegraphics[scale=0.8]{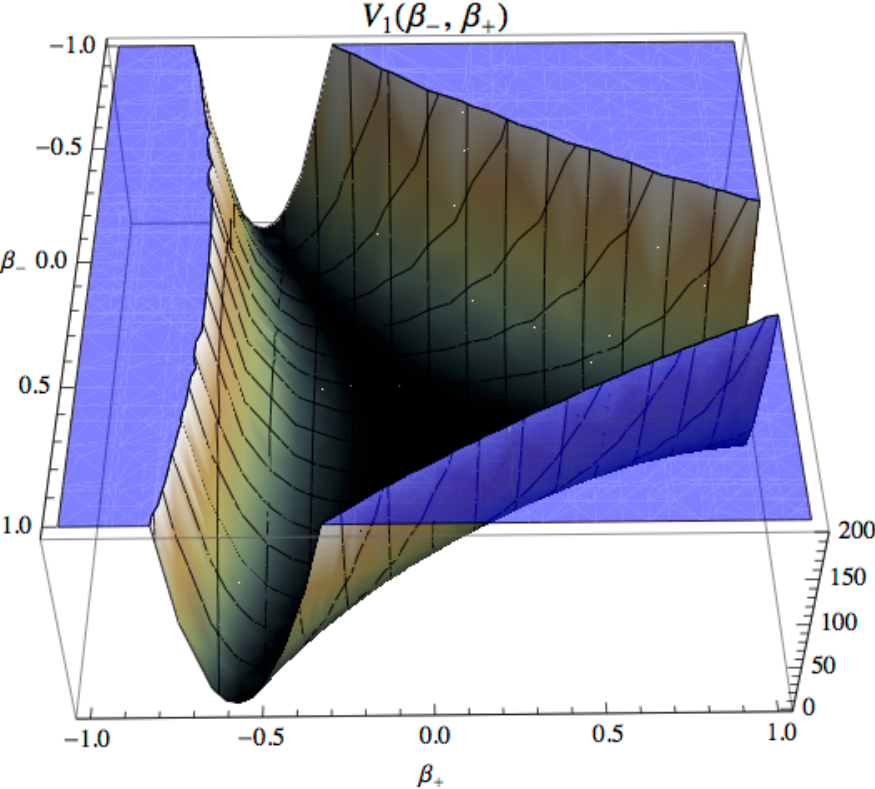}
\caption{The plot of $V_\mfn $ for $\mfn=1$ near its minimum. Boundedness from
below,  confining aspects, and $C_{3v}$ symmetry are illustrated.}
\label{figure2}
\end{figure}

The potential $W_{\mfn}$ is {\it bounded} from below and reaches
its minimum value, $W_{\mfn}=- \mfn^2$, at
$\beta_\pm=0$. The potential $W_{\mfn}$ is expanded around its minimum as follows:
\begin{equation}\label{potharmonic}
W_{\mfn}(\beta) = - \mfn^2 + 8
\mfn^2(\beta_+^2+\beta_-^2) +o(\beta_\pm^2)\,,
\end{equation}
which clearly shows its two-dimensional harmonic approximation.
The $W_{\mfn}$ is asymptotically {\it confining} except for the following three
directions (shown in  Fig.\;\ref{figure1}), in which $W_{\mfn} \to 0$: $(i) ~\beta_-=0, ~\beta_+ \to +\infty,~ $ $(ii)~\beta_+=- \frac{\beta_-}{\sqrt{3}}, ~\beta_- \to +\infty,~$ $(iii)~\beta_+=  \frac{\beta_-}{\sqrt{3}}, ~\beta_- \to -\infty$.\\
The form of the Bianchi-IX potential deserves particular attention
due to its three ``open'' ${\sf C}_{3v}$ symmetry directions (see
figures \ref{figure1} and \ref{figure2}). One can view them as three
deep ``canyons'', increasingly
narrow until  their respective wall edges close up at the infinity
whereas their respective bottoms tend to zero. The motion of the
Misner particle in this potential is chaotic \cite{RS}. Though the
curvature, which is proportional to the potential, flattens with
time, the confined particle undergoes
infinitely many oscillations. In the so-called steep wall
approximation, the particle is locked in the triangular potential
with its infinitely steep walls moving apart in time. At the
quantum level, the confining shape originates a discrete spectrum.
On the other hand, it is unclear on a mathematical level whether or not the Bianchi-IX
potential also originates a continuum spectrum.
Nevertheless the following physical reasoning leads to the conclusion that a continuous spectrum probably does not exist. Indeed a continuous spectrum should be associated with the eigensystem $\{ (\psi_k(\beta_-,\beta_+), \epsilon_k) \}_{k \in \mathcal{D} \subset \mathbb{R}^2}$ where: (i) the  scattering eigenstates (not $\mathbb{R}^2$-Lebesgue square integrable) $\psi_k(\beta)$ represent the possibility for the system to go from infinity to infinity in the $(\beta_-,\beta_+)$-plane, (ii) $k \in \mathcal{D} \subset \mathbb{R}^2$ represents the $2D$-momentum of the incoming ``particles", (iii) $k$ belongs to some $2D$-domain $\mathcal{D}$ of $\mathbb{R}^2$ ($\mathcal{D}$ is not restricted to a subset of measure $0$ for the $\mathbb{R}^2$-Lebesgue measure), (iv) $k \in \mathcal{D}$ leads to an interval (generally unbounded) of eigenvalues $\epsilon_k$ (the continuous spectrum). In our case the potential only possesses three discrete ``open" directions at infinity, therefore the incoming momentum $k$ cannot belong to a $2D$-domain $\mathcal{D}$ (a subset with a non-vanishing $\mathbb{R}^2$-Lebesgue measure). Then it is reasonable to think that a continuous spectrum does not exist. 

The evolution of Bianchi IX can be viewed as a non-linear model of
 gravitational wave in dynamical isotropic geometry (see e.g.
\cite{cwm2,King} and references therein). The wave, which consists
of two nonlinearly coupled components $\beta_{\pm}$, is
homogenous, that is, it is a pure time oscillation. Its wavelength
is thought to be much larger than the extension of the considered
patch of the universe with its spatial derivatives neglected.  The
energy of the wave sources the gravitational contraction. As we
later show, the quantization of the wave introduces important
modifications to the dynamics of the whole universe. The
qualitative study of spatially homogeneous models was pioneered by
Bogoyavlensky (see \cite{Bogoyavlensky1985} and references
therein).

\subsection{Redefinition of basic variables}

For the purpose of ACS quantization, we  redefine the isotropic phase
space variables as the canonical pair:
\begin{equation}
q :=e^{3\beta_0/2}, ~~~p := \frac{2}{3}e^{-3\beta_0/2}p_0~.
\end{equation}
Note that $(q,p)$ lives in the half-plane. The Hamiltonian \eqref{Hamilphys0}
now assumes the form
\begin{equation}\label{Hamilphys1}
\mathsf{H}= -\frac{\mathcal{N}}{24} \left(\frac{9}{4} \, p^2 -
\frac{ p_+^2+p_-^2}{q^2}-36 q^{2/3} W_{\mfn}(\beta) \right) \,.
\end{equation}
Let us split the potential $W_{\mfn}$ into its isotropic and anisotropic
components. The isotropic part corresponds to that part of curvature
which is independent of $\beta_{\pm}$, whereas the anisotropic part
vanishes for $\beta_{\pm}=0$:
\begin{equation}\label{ani-potential}
W_{\mfn}(\beta)= - \mfn^2+V_{\mfn}(\beta)
\end{equation}

The lapse function is not dynamical and its choice is irrelevant for
the classical dynamics as it only fixes the magnitude and direction of
the Hamiltonian flow in the constraint surface. From now on we put the
lapse $\mathcal{N}=-24$. The Hamiltonian \eqref{Hamilphys1} reads now
\begin{equation}\label{Hamilphys2}
\mathsf{H}= \frac{9}{4} \, p^2+36\mfn^2q^{2/3} -\mathsf{H}_q \,,
\end{equation}
where $\mathsf{H}_q$ is the $q$-dependent Hamiltonian for the anisotropic variables,
\begin{equation}
\label{Hamilphys2pm}
\mathsf{H}_q :=\frac{ p_+^2+p_-^2}{q^2}+36 q^{2/3} V_{\mfn}(\beta) \,.
\end{equation}

\subsection{Discussion of the constraint}

The analytical expressions \eqref{Hamilphys2}-\eqref{Hamilphys2pm}
for $\mathsf{H}$  remind us of molecular system's Hamiltonian. The
pair $(q,p)$ plays the role of the nucleus-like dynamical
variables and $(\beta_\pm,p_\pm)$ are electron-like dynamical
variables. Quantum molecular systems are usually considered by
making use of the Born-Oppenheimer approximation (BO) or its
``diagonal correction'' named Born-Huang  (BH)
\cite{born51,sutcliffe12}. In molecular physics, the validity of
these approximations depends crucially on the ratio between nuclei
and electron masses. Namely, a nucleus mass is very large when
compared to the electron mass. In our case, Eq.\;\eqref{Hamilphys2pm} indicates that $q^2$ plays a role of  ``mass'' for the anisotropic variables, whereas Eq.\;\eqref{Hamilphys2} shows that the ``mass'' of the isotropic variable is constant.
Therefore, near the singularity $q=0$, $\beta_\pm$'s become \emph{light} degrees of freedom and $q$ with its constant mass may be treated as a \emph{heavy} degree of freedom. Hence, we may follow either the BO or the BH approximation scheme in quantizing our system. This issue is considered in more detail in the next
section.

For Eqs.\;\eqref{Hamilphys2}-\eqref{Hamilphys2pm}, one checks that
$\beta_\pm = 0 = p_\pm $ is a solution to the Hamilton equations
of motion. In this case, the constraint $\mathsf{H}=0$ reduces  to
\begin{equation}
\label{classconstraint} \frac{9}{4} \, p^2+36\mfn^2q^{2/3} =0 \,.
\end{equation}
and we recover the closed vacuum FRW constraint, which possesses
the unique singular and uninteresting solution $p=0=q$. Nevertheless,
it makes sense to consider a small perturbation $\delta \beta_\pm$ from
$\beta_\pm=0$. The dynamical equation for $\delta \beta_\pm$ based on
the harmonic approximation of Eq.\;\eqref{potharmonic}) is
\begin{equation}\label{waves}
\delta \ddot{\beta}_\pm  = -2 \frac{\dot{q}}{q} \delta
\dot{\beta}_\pm - 2 \frac{\mfn^2}{q^{4/3}} \delta \beta_\pm
\,.
\end{equation}
It can be demonstrated that the Friedmann model evolving towards the
singularity is not stable in the phase space of the Bianchi IX
model. More precisely, any perturbation of isotropy will grow and
develop into an oscillatory and chaotic behavior. The growth of
isotropy is apparent from Eq.\;\eqref{waves} for $\dot{q}/q<0$.
As the shear grows, the harmonic approximation breaks down and
fully non-linear dynamics develops. As we show, this
behavior  is suppressed on the quantum level and allows the
harmonic approximation to remain valid all the way towards the
big bounce. Moreover, we show in Section \ref{consistency} that
the BO approximation also survives the bounce phase. As the shear
is known to dominate over any type of familiar matter close
enough to the singularity, we consider only the vacuum case.
Then the lack of solutions for the vacuum isotropic universe
as concluded from \eqref{classconstraint} leads to the prediction
that one should take into account the effect of the `quantum zero
point energy' generated by the quantized anisotropy degrees
of freedom of the Bianchi-IX model.

\section{Quantization}

\label{quantiz} In what follows we apply a quantization based on
the Dirac method and inspired by Klauder's approach
\cite{klauderscm}: (i) quantizing $\mathsf{H}$  in kinematical
phase space, (ii) finding the semi-classical expression
$\check{\mathsf{H}}$ of the quantum Hamiltonian $\hat{\mathsf{H}}$
using Klauder's approach and our adiabatic approximation, and
(iii) implementing the Hamiltonian constraint on the
semi-classical level
$\check{\mathsf{H}} = 0$.\\

\subsection{Quantum constraint}

Since in Eqs.\;\eqref{Hamilphys2}-\eqref{Hamilphys2pm} we have
$(q,p) \in \mathbb{R}^+_{\ast} \times \mathbb{R}$ and $(\beta_\pm, p_\pm)
\in \mathbb{R}^4$, we can follow the idea of our previous paper
\cite{Bergeron:2013ika} in the realization of the step (i):
\begin{itemize}
   \item[(i)] for the quantization of functions (or distributions) of the pair $(q,p)$
   living in the open half-plane,
   we apply ACS
   quantization, whose principles and methods, as part of integral
   quantizations (see \cite{bergaz14} and references therein), are explained in
   \cite{Bergeron:2013ika}
   and summarized in
   Appendix \ref{affquant}. This procedure yields $\hat{p}=-i \hbar
   \partial_x$ and $\hat{q}$ defined as the multiplication by $x$,
   both acting in the Hilbert space $L^2(\mathbb{R}^+, \ud x)$,
   \item[(ii)] for quantizing functions of the pairs $(\beta_\pm, p_\pm)$ we can
   choose either the Weyl-Heisenberg coherent state (CS) quantization \cite{bergaz14},
   which has a regularizing r\^ole, or, directly, the  canonical
   quantization. Due to the simplicity of the model we follow the later
   option. Actually both yield $\hat{p}_\pm = - i
   \hbar\partial_{\beta_\pm}$ and the multiplication operator
   $\hat{\beta}_\pm$, both acting in $L^2(\mathbb{R}^2, \ud \beta_+
   \ud \beta_-)$.
\end{itemize}
Thus, for the quantum Hamiltonian $\hat{\mathsf{H}}$ corresponding
to Eqs.\;\eqref{Hamilphys2}-\eqref{Hamilphys2pm} we get
\begin{equation}
\label{quham}
 \hat{\mathsf{H}}=\frac{9}{4} \left(
 \hat{p}^2 + \frac{\hbar^2 \frak{K}_1}{\hat{q}^2} \right) + 36\mfn^2\frak{K}_3 \hat{q}^{2/3} -
 \hat{\mathsf{H}}_{\hat{q}}\,,
\end{equation}
\begin{equation}
\label{eq:hamilpm}
\hat{\mathsf{H}}_q :=\frak{K}_2\frac{\hat{p}_+^2+\hat{p}_-^2}{q^2}+
36 \frak{K}_3 q^{2/3}V_{\mfn}(\beta)\,.
\end{equation}
where the $\frak{K}_i$ are positive and non-vanishing purely numerical constants dependent on
the choice of the so-called {\it fiducial} vector. With the choice
made in our previous paper \cite{Bergeron:2013ika}, and thanks to the
formula recalled in Appendix \ref{affquant}, we have
\begin{align}
\nonumber \frak{K}_1 = &\frac{1}{4} \left( 1+ \nu \frac{K_0(\nu)}{K_1(\nu)}
\right),  \quad \frak{K}_2 =\left(
\frac{K_2(\nu)}{K_1(\nu)}\right)^2\,, \\
 \label{Ki}& \frak{K}_3
=\frac{K_{5/3}(\nu)}{K_1(\nu)^{1/3} K_2(\nu)^{2/3}} \,,
\end{align}
where $\nu >0$ is a free parameter and the $K_r(\nu)$ are
the modified Bessel functions \cite{magnus66}. Since we deal with ratios
of such functions throughout the sequel, we adopt the convenient notation
\begin{equation}
\label{xibessel}
\xi_{rs} = \xi_{rs}(\nu)=  \frac{K_r(\nu)}{K_s(\nu)}= \frac{1}{\xi_{sr}}\, .
\end{equation}
One convenient feature of such a notation is that $\xi_{rs}(\nu) \sim 1$
as $\nu \to \infty$ (a consequence of  $K_r(\nu) \sim \sqrt{\pi/(2\nu)}$).
Thus \eqref{Ki} reads
\begin{align}
\nonumber \frak{K}_1 & = \frac{1}{4} \left( 1+ \nu\, \xi_{01}(\nu)
\right)\, , \quad \frak{K}_2 =\left(
 \xi_{21}(\nu)\right)^2\, ,\\
\label{xirs}  & \frak{K}_3
= \left( \xi_{\frac{5}{3} 1}\right)^{1/3}\,\left( \xi_{\frac{5}{3} 2}\right)^{2/3}
\end{align}

There exist  many other choices of fiducial vectors yielding similar
constants $\frak{K}_i$.
These vectors depend themselves on arbitrary parameters which can
be suitably adjusted. Also, the Hamiltonian \eqref{quham} itself
is defined up to a multiplicative factor. It is also crucial to recall that there exists an infinite range of values for constants $\frak{K}_i$, for which
the symmetric $\hat{\mathsf{H}}$ has a unique self-adjoint extension.
This is proved by making use of the reasoning previously presented
in \cite{Bergeron:2013ika} and recalled in Appendix \ref{affquant}. 

Important remark: The analysis developed in the following sections is independent of the specific numerical values given to the coefficients $\frak{K}_i$ as long as the bounce happens for small enough values of $q$ in order for the adiabatic approximation to be valid (see below). In other words, all the following results determine the qualitative features of the dynamics unambiguously and the numerical coefficients pin down the dynamics quantitatively.\\

\subsection{Semi-classical Lagrangian approach}
\label{appsclag}

Being inspired by Klauder's approach \cite{klauderscm}, we present
a consistent framework allowing to approximate the quantum
Hamiltonian and its associated dynamics (in the constraint surface) by
making use of the semi-classical Lagrangian approach.

\subsubsection{General setting}

The quantum Hamiltonian \eqref{quham} has the general form (up to
constant factors)
\begin{equation}
\label{quhamgen}
 \hat{\mathsf{H}}=\mathcal{N}\left( \hat{p}^2 + \frac{K}{\hat{q}^2}+L \hat{q}^{\frac{2}{3}}
 - \hat{\mathsf{H}}^{(\mathrm{int})}( \hat{q})\right) \,,
\end{equation}
where $K$ and $L$ are some positive constants and the $ q$-dependent
Hamiltonian $\hat{\mathsf{H}}^{(\mathrm{int})}(q)$ (also denoted by $\hat{\mathsf{H}}_q$
in the previous subsection) acts on Hilbert
space of states for `internal' degrees of freedom, i.e., the
anisotropic ones.

The Schr\"odinger equation, $i \hbar \frac{\partial }{\partial t}
| \Psi(t) \rg = \mathcal{N}\hat{\mathsf{H}} | \Psi(t) \rg$, can be deduced
from the Lagrangian:
\begin{equation}
\label{lagrangen} {\sf L} (\Psi, \dot \Psi, \mathcal{N}):=\lg
\Psi(t) |\left( i\hbar\frac{\partial}{\partial t}
-\mathcal{N}\hat{\mathsf{H}}\right) |\Psi(t)\rg \, ,
\end{equation}
via the variational principle with respect to $|\Psi(t)\rg$. In order to solve the Schr\"odinger equation, the lapse function $\mathcal{N}$ should be fixed.

The quantum counterpart  of the classical constraint
$\mathsf{H}=0$ can be obtained as follows
\begin{equation}
\label{vmH0}
\dfrac{\partial {\sf L}}{\partial  \mathcal{N}} = \lg
\Psi(t) |\hat{p}^2 + \frac{K}{\hat{q}^2}+L\hat{q}^{\frac{2}{3}}-
\hat{\mathsf{H}}^{(\mathrm{int})}( \hat{q})|\Psi(t)\rg=0\, .
\end{equation}
The commonly used Dirac's way of imposing a constraint,
$\hat{\mathsf{H}}|\Psi(t)\rg = 0$, implies \eqref{vmH0} but the
reciprocal does not hold in general.

At this stage, we suppose (due to the confining character of the
potential $V_{\mfn}$) that there exists
$\hat{\mathsf{H}}^{(\mathrm{int})} (q)$ as self-adjoint operator
(and as a function of the c-number $q$) with purely discrete
spectral decomposition, which is of course true in its harmonic approximation,
\begin{equation}
\label{specdecinq} \hat{\mathsf{H}}^{(\mathrm{int})}(q) = \sum_n
E^{(\mathrm{int})}_n(q) \,
|\phi^{(\mathrm{int})}_n\rg\lg\phi^{(\mathrm{int})}_n|\,.
\end{equation}

To present the Klauder semi-classical procedure
 in the most general situation (not restricted to
Bianchi-IX), we distinguish the two cases:
\begin{enumerate}
  \item[(i)]   $\phi^{(\mathrm{int})}_n$ is {\it independent} of $q$, which
  allows a complete separation of variables, and leads to the original
  Born-Oppenheimer \cite{born51,sutcliffe12} approach;
  \item[(ii)]   $\phi^{(\mathrm{int})}_n$ is {\it dependent} on $q$.
\end{enumerate}

Different semi-classical approximations result depending on these
choices\footnote{We first present the case (i), which is simple,
and later introduce the more complicated case (ii), being applied
to the Bianchi IX model.}.

\subsubsection{Semi-classical Lagrangian approximations \label{semiclasslagrange}}

{\bf (i) $\phi^{(\mathrm{int})}_n$ independent of $q$}

In this case, a family of exact solutions of the time-dependent
Schr\"odinger equation $i\hbar\frac{\partial}{\partial t}
|\Psi(t)\rg = \hat{\mathsf{H}} |\Psi(t)\rg $ can be introduced in
the form of the tensor product
\begin{equation}
\label{sepvartens} |\Psi(t)\rg = |\phi(t)\rg
\otimes|\phi^{(\mathrm{int})}_n\rg\,,
\end{equation}
where $|\phi(t)\rg$ is solution to the reduced time-dependent
Schr\"odinger equation
\begin{align}
\label{tdepredSE}
i & \hbar\frac{\partial}{\partial t} |\phi(t)\rg
=\\
\nonumber & \mathcal{N}\left(\hat p^2 + \frac{K}{\hat{q}^2}+L\hat{q}^{\frac{2}{3}} -
E^{(\mathrm{int})}_n( \hat{q}) \right)  |\phi(t)\rg =: \\
\nonumber & \mathcal{N}
\hat{\mathsf{H}}^{\mathrm{red}}_n |\phi(t)\rg  \,
\end{align}
where $E^{(\mathrm{int})}_n$ is the eigenvalue of
$\hat{\mathsf{H}}^{(in)}$. The tensor product $|\Psi(t)\rg $ of Eq.\;\eqref{sepvartens}
is precisely the  Born-Oppenheimer-like  solution \cite{combes80}. The
equation \eqref{tdepredSE} may be derived from a variational
principle applied to the quantum Lagrangian
\begin{align}
\label{lagrangian(i)} {\sf L}^{\mathrm{red}} (\phi, \dot \phi,
\mathcal{N})& :=\\
\nonumber & \lg \phi(t) | \left(i\hbar\frac{\partial}{\partial
t}
 - \mathcal{N} \hat{\mathsf{H}}^{\mathrm{red}}_n\right) |\phi(t)\rg \, .
\end{align}
Following Klauder \cite{klauderscm}, we assume that  $|\phi(t)\rg$
is in fact an affine coherent state. We assume in the following
that the fiducial vector $\psi$ has been chosen such that
$c_0(\psi)=c_{-1}(\psi)$ in order to obtain the canonical rule
$[A_q,A_p]=i \hbar$. Furthermore, we need to apply a rescaling $|
q(t),p(t)\rg \to | \lambda q(t),p(t)\rg $ in order to ensure $\lg
\lambda q(t), p(t) |\,A_q \,|q(t),p(t)\rg = q(t)$ and $\lg \lambda
q(t), p(t) |\, A_p \,|q(t),p(t)\rg = p(t)$.  The parameter
$\lambda$ is uniquely defined by the choice of the fiducial
vector, namely $\lambda=1/c_{-3}(\psi)$ (see appendix \ref{affquant}).

Therefore we
replace $| \Psi(t) \rg$ in ${\sf L}$ of Eq.\;\eqref{lagrangen} by
\begin{equation}
|  \Psi(t) \rg = |\lambda q(t), p(t) \rg \otimes | \phi^{(\mathrm{int})}_n
\rg \,.
\end{equation}
where $q(t)$ and $p(t)$ are some time-dependent functions. Then
the Lagrangian  \eqref{lagrangen}  or \eqref{lagrangian(i)} turns
to assume the semi-classical form
\begin{align}
\label{lagrangian(i)sc}  {\sf L}^{\mathrm{sc}} (q,\dot q,
p, \dot p, \mathcal{N}) &=\\
 \nonumber \lg \lambda q(t),p(t) | \left(
i\hbar\frac{\partial}{\partial t} -
\mathcal{N}\hat{\mathsf{H}}^{\mathrm{red}}_n
\right) &| \lambda q(t),p(t)\rg \\ 
\nonumber = - q\dot p -
\mathcal{N}\lg \lambda q(t),p(t)|&  \hat{\mathsf{H}}^{\mathrm{red}}_n
| \lambda q(t),p(t)\rg \\
\nonumber = - \frac{d}{dt}(qp) + \dot{q} p\\
\nonumber  -
\mathcal{N}\lg \lambda q(t),p(t)| &\hat{\mathsf{H}}^{\mathrm{red}}_n
| \lambda q(t),p(t)\rg\, .
\end{align}
The appearance of the first term $- q\dot p $ in the r.h.s.
of this equation results from the derivative  of  \eqref{affrep+}
with respect to parameters $q$ and $p$ leading to  \eqref{dercs}.

The semi-classical expression for the Hamiltonian is the lower
symbol
\begin{equation}\label{sce}
\check{\mathsf{H}}^{\mathrm{red}}_n (q,p) := \lg \lambda q,p|
\hat{\mathsf{H}}^{\mathrm{red}}_n | \lambda q,p\rg\, .
\end{equation}
It is defined by the `frozen' quantum eigenstate `$n$' of the
internal degrees of freedom.

From this reduced Hamiltonian one derives the equations of motion
together with the constraint
\begin{align}
\label{eqmotred}
  \dot q   & = \mathcal{N} \frac{\partial}{\partial p}
  \check{\mathsf{H}}^{\mathrm{red}}_n (q,p),\\
   \dot p &= -  \mathcal{N} \frac{\partial}{\partial q}
   \check{\mathsf{H}}^{\mathrm{red}}_n (q,p) \, \\
  0  &= \check{\mathsf{H}}^{\mathrm{red}}_n (q,p) \,.
\end{align}
These equations will allow us to set up Friedmann-like equations
with quantum corrections for $q$ and $p$.

{\bf (ii)  $\phi^{(\mathrm{int})}_n$ dependent of $q$}

Let us examine the general case in which the eigenstates
$|\phi^{(\mathrm{int})}_n\rg$ depend on $q$. We start again from
the spectral decomposition
\begin{equation}
\label{specdecinqq} \hat{\mathsf{H}}^{(\mathrm{int})}(q) = \sum_n
E^{(\mathrm{int})}_n(q) \,
|\phi^{(\mathrm{int})}_n(q)\rg\lg\phi^{(\mathrm{int})}_n(q)|\,,
\end{equation}
and we pick some other $q$-independent orthonormal basis
$|e^{(\mathrm{int})}_n\rg$ of the `internal'  Hilbert space
$\mathcal{H}^{(\mathrm{int})}$. This change of basis is associated
with the introduction of the $q$-dependent unitary operator
\begin{equation}
\label{Unchbas} U(q):= \sum_n |\phi^{(\mathrm{int})}_n(q)\rg\lg
e^{(\mathrm{int})}_n|\,,
\end{equation}
which allows to deal with the analogue of the Hamiltonian
\eqref{specdecinq}:
\begin{align}
\label{specdecinen} \widetilde{\mathsf{H}}^{(\mathrm{int})}(q) = &
U^\dag(q)\hat{\mathsf{H}}^{(\mathrm{int})}(q) U(q) =\\
\nonumber & \sum_n
E^{(\mathrm{int})}_n(q) \, |e^{(\mathrm{int})}_n\rg\lg
e^{(\mathrm{int})}_n|\,.
\end{align}
The quantum Hamiltonian \eqref{quhamgen} has now the general form
\begin{equation}
\label{quhamgentild}
 \hat{\mathsf{H}}=\mathcal{N} \left( \hat{p}^2 + \frac{K}{\hat{q}^2}
 +L\hat{q}^{\frac{2}{3}}- U(\hat q)\widetilde{\mathsf{H}}^{(\mathrm{int})}(\hat q)U^{\dag}
 (\hat q)\right) \,.
\end{equation}

The difference between the Hamiltonians of cases \textbf{(i)} and \textbf{(ii)} is
the presence in \textbf{(ii)} of the unitary operator $U(\hat{q})$ that
introduces a quantum correlation (entanglement) between the
`internal' degrees of freedom (anisotropy) and the `external' one
(the scale factor). As a consequence, any solution $|\Psi(t)\rg$
of the time-dependent Schr\"odinger equation
$i\hbar\frac{\partial}{\partial t} |\Psi(t)\rg = \hat{\mathsf{H}}
|\Psi(t)\rg $ cannot be factorized as a tensor product like
$|\phi(t)\rg \otimes|\phi^{(\mathrm{int})}(t)\rg$, contrarily to
the case (i). In our case we wish to follow Klauder's approach to
build some semi-classical Lagrangian analoguous to
Eq.\;\eqref{lagrangian(i)}. We use the previous case (i) as a
starting point (as a guide) to build approximate possible forms of
$|\Psi(t)\rg$.\\

It is interesting to notice that the Hamiltonian
$\hat{\mathsf{H}}$ of Eq.\;\eqref{quhamgentild} is unitarily
equivalent to the one that occurs in quantum electrodynamics. For
this purpose let us introduce the $q$-dependent operator
$\hat{\mathrm{A}}(q)$ acting on the Hilbert space
$\mathcal{H}^{(\mathrm{int})}$ of internal degrees of freedom as
\begin{equation}
\label{gaugefield} \hat{\mathrm{A}}(q) = i \hbar \frac{\ud U}{\ud
q}(q) U^\dag (q) \,.
\end{equation}
As a matter of fact, $\hat{\mathrm{A}}(q)$ is self-adjoint, and
the Hamiltonian $\hat{\mathsf{H}}$ of Eq.\;\eqref{quhamgentild}
reads as
\begin{align}
\label{quhamgentransform}
& \hat{\mathsf{H}} = \mathcal{N}  \,
U(\hat q) \times \\
\nonumber  & \left( (\hat{p} - \hat{\mathrm{A}}(\hat{q}))^2 +
\frac{K}{\hat{q}^2} +L\hat{q}^{\frac{2}{3}}- \widetilde{\mathsf{H}}^{(\mathrm{int})}(\hat
q)\right) U^{\dag}(\hat q) \,.
\end{align}
The form \eqref{quhamgentransform} (i.e. modulo the unitary transformation $U(\hat q)$) is the most general form of the Hamiltonian of a particle moving on a half-line and minimally coupled to an external field. This form is also most general on the grounds of the so called shadow principle \cite{JMLL}. According to the principle,  at any fixed time the state of an interacting particle should be indistinguishable from the state of a free particle and the Galilean addition rule of velocities should be preserved. In particular, if we interpret $\widetilde{\mathsf{H}}^{(\mathrm{int})}(q)$ as a $q$-dependent electromagnetic energy (despite all the differences) and $\hat{\mathrm{A}}(q)$ as a gauge field, the problem appears (up to a unitary transformation) similar to the one of a charged
particle in interaction with an electromagnetic field.\\

Now, using   \eqref{quhamgentild} and taking into account the
analysis of the previous case \textbf{(i)}, one can define different
possible expressions of $| \Psi(t) \rg$:

\begin{enumerate}
  \item[(a)] In the first approach we keep the tensor product expression
  of \textbf{(i)}, but inserting the $q$-dependence of eigenstates. This corresponds
to
  a Born-Oppenheimer-like approximation:
  \begin{equation}
\label{firstapp} |\Psi(t)\rg \approx | \lambda q(t),p(t)\rg\otimes
|\phi^{(\mathrm{int})}_n(q(t))\rg \,.
\end{equation}
  \item[(b)] The second strategy consists in  introducing some (minimal)
  entanglement between $q$ and `internal' (anisotropy) degrees of freedom.
  This corresponds to a Born-Huang-like approximation:
   \begin{equation}
\label{secapp} |\Psi(t)\rg \approx U(\hat q)
\left(| \lambda q(t),p(t)\rg\otimes  |e^{(\mathrm{int})}_n\rg\right)\,.
\end{equation}
  \item[(c)] In the third method one  keeps the tensor product approximation,
  but including a general time-dependent state for the internal degrees of
  freedom:
   \begin{equation}
\label{thirdapp} |\Psi(t)\rg \approx | \lambda q(t),p(t)\rg\otimes
|\phi^{(\mathrm{int})}(t)\rg \,.
\end{equation}
  \item[(d)] The fourth strategy is the most  general one.  It  consists
  in merging (b) and
  (c):
   \begin{equation}
\label{fourthapp} |\Psi(t)\rg \approx U(\hat q) \left(
|\lambda q(t),p(t)\rg\otimes  |\phi^{(\mathrm{int})}(t)\rg \right) \,.
\end{equation}
\end{enumerate}

Building now the semi-classical Lagrangian in agreement with the
procedure defined in \textbf{(i)}, we can distinguish two categories in the
approximations listed above.
\begin{itemize}
   \item[(1)] (a) and (b) are completely manageable on the
    semi-classical level: they involve $q$ and $p$ as dynamical
    variables, while the anisotropy degrees of freedom are `frozen' in
    some eigenstate; (a) and (b) correspond typically to adiabatic
    approximations.
\item[(2)] (c) and (d) are more complicated: they
mix a semi-classical dynamics for $(q,p)$ and a quantum dynamics
for the anisotropy degrees of freedom. This corresponds to
`vibronic-like' approximations,  well-known in molecular physics and quantum chemistry \cite{yakorny12}. In
our case this means that different quantum eigenstates of the
anisotropy degrees of freedom are involved in the dynamics: during
the evolution excitations and decays are possible, with an
exchange of energy with  the `classical degree of freedom'
$(q,p)$.
\end{itemize}

The Bianchi-IX Hamiltonian belongs to the general case \textbf{(ii)}. So
the different approximations (a), (b), (c), (d),  presented above can
be tested. In this paper we restrict ourselves to the presentation
of the simplest cases (a) and (b). We postpone the study of the
more complicated cases (c) and (d) to future papers.

\subsection{Semiclassical constraint}

We now proceed to the spectral analysis  of the operator
$\hat{\mathsf{H}}$ by making use of BO-like and BH-like
approximations, presented in a general form in subsection \ref{appsclag}, and to
its semi-classical analysis through affine coherent states.

\subsubsection{Born-Oppenheimer approximation}
\label{BOap}

In this approximation we assume that the anisotropy degrees of
freedom are frozen in some eigenstate $
|\phi^{(\mathrm{int})}_n(q(t))\rg$, evolving adiabatically,  of the
$q$-dependent Hamiltonian $\hat{\mathsf{H}}_q$ given by \eqref{eq:hamilpm}.
If we denote by $E_{N}(q)$ the
eigenenergies of $\hat{\mathsf{H}}_q$, the reduced
Hamiltonian $\hat{\mathsf{H}}^{\mathrm{red}}_N$ of
Eq.\;\eqref{tdepredSE} reads
\begin{equation}
\label{bornoppenhamred} \hat{\mathsf{H}}^{\mathrm{red}}_N =
\frac{9}{4} \left(
 \hat{p}^2 + \frac{\hbar^2 \frak{K}_1}{\hat{q}^2} \right) + 36\mfn^2\frak{K}_3\hat{q}^{2/3} - E_{N}(\hat{q}) \,.
\end{equation}
Due to the harmonic behavior of $V_{\mfn}$ near its minimum, i.e.,
\begin{equation}
V_{\mfn}(\beta) = 8 \mfn^2
(\beta_+^2+\beta_-^2) + o(\beta_\pm^2) \,,
\end{equation}
the harmonic approximation to the eigenenergies $E_N(q)$
is manageable ($N=0,1,\dots$), giving
\begin{equation}
E_N(q) \simeq \frac{24 \hbar}{q^{2/3}} \mfn \sqrt{2 \frak{K}_2
\frak{K}_3} \, (N+1) \,.
\end{equation}
In fact $N=n_++n_-$, $n_{\pm}\in \N$. More details, including a discussion of another common approximation made for the anisotropy potential, are given in
Appendix \ref{appanqh}.\\

\noindent {\sf Remark}: The harmonic form of  $E_N(q)$ is
an increasingly rough approximation for large values of $N$,
since $V_{\mfn}$ is highly non-harmonic far from its minimum. But for
small values of $N$, this expression is valid for any value of
$q$. The steep wall approximation (see Appendix \ref{appanqh}) is
able to give a better expression for the eigenenergies as their values go
to infinity. However, both approximations do not change the main line
of reasoning in what follows.\\

Taking into account the rescaling of affine coherent states (see
Sec. \ref{semiclasslagrange}), the semi-classical expression
$\check{\mathsf{H}}^{\mathrm{red}}_N$ involved in Eq.\;\eqref{sce}
is defined as
\begin{equation}\label{lam}
\check{\mathsf{H}}^{\mathrm{red}}_N(q,p) = \langle \lambda q, p |
\hat{\mathsf{H}}^{\mathrm{red}}_N | \lambda q, p \rangle \, ,
\end{equation}
where $\lambda := \xi_{02}(\nu)$ is chosen to get the exact
correspondence (see
Sec. \ref{semiclasslagrange})
\begin{equation}
\label{excorr}
\langle
\lambda q, p | \hat{q}| \lambda q, p \rangle = q\, , \quad   \langle
\lambda q, p | \hat{p}| \lambda q, p \rangle = p\,.
\end{equation}
Finally, we obtain
\begin{align}
\label{bornoppensemi} \check{\mathsf{H}}^{\mathrm{red}}_N(q,p)  &=
\frac{9}{4} \left( p^2 + \frac{\hbar^2 \frak{K}_4}{q^2} \right) +\\
\nonumber & 36 \mfn^2
\frak{K}_5 q^{2/3} - \frac{24 \hbar}{q^{2/3}} \frak{K}_6
\mfn (N+1) \,,
\end{align}
where the three new constants $ \frak{K}_i=  \frak{K}_i(\nu)$ are given by
\begin{equation}
\label{consK456}
\begin{split}
 \frak{K}_4 =  (\xi_{10})^2\, &\left( \frac{\nu^2}{16}+\frac{\nu}{4} \xi_{21}+\frac{3\nu}{8} \xi_{10}+\xi_{20} \right)\, ,\\
  \frak{K}_5=& \xi_{\frac{2}{3}1}\, \left(\xi_{\frac{5}{3}0}\right)^{1/3}
 \left(\xi_{\frac{5}{3}2}\right)^{2/3}\, , \\
  \frak{K}_6 =& \sqrt{2}\,
 \left( \xi_{10} \right)^{4/3} \,  \left(\xi_{20}\right)^{5/3} \left(\xi_{\frac{5}{3}1}\right)^{1/2}\, .
\end{split}
\end{equation}
For large values of $\nu$ (typically $\nu \gtrsim 20$) we get
\begin{equation}
\frak{K}_4 \simeq \frac{\nu^2}{16},~~~
 \frak{K}_5 \simeq 1,~~~
 \frak{K}_6 \simeq \sqrt{2}\; .
\end{equation}

\subsubsection{Born-Huang approximation}
\label{BHap}
In the Born-Huang approximation framework (see
Sec. \ref{semiclasslagrange}), we also assume that the anisotropy
degrees of freedom are frozen in some  eigenstate, but it is an eigenstate
$|e^{(\mathrm{int})}_n\rg$ of the Hamiltonian $
\widetilde{\mathsf{H}}^{(\mathrm{int})}(q)$ introduced in Eq.\;\eqref{specdecinen}.
In the case of
Bianchi-IX, thanks to the harmonic approximation of $V_{\mfn}(\beta)$, we
can find an approximation for $
\widetilde{\mathsf{H}}^{(\mathrm{int})}(q)$  and the corresponding unitary
operator $U(q)$
in Eq.\;\eqref{Unchbas}.
We get
\begin{equation}
U(q) \simeq e^{\frac{2 i}{3} (\ln q) \, \hat{\mathrm{D}}}
\,,
\end{equation}
with
\begin{equation}
\hat{\mathrm{D}} = \hat{\mathrm{D}}_+ + \hat{\mathrm{D}}_- \, ,
\quad \hat{\mathrm{D}}_\pm = \frac{1}{2 \hbar}(\hat{p}_\pm
\hat{\beta}_\pm + \hat{\beta}_\pm \hat{p}_\pm) \,.
\end{equation}
We deduce for the `gauge field' $\hat{\mathrm{A}}(q)$, the expression
in Eq.\;\eqref{gaugefield}, becomes
\begin{equation}
\hat{\mathrm{A}}(q) = - \frac{2 \hbar}{3 q} \hat{\mathrm{D}} \,.
\end{equation}
As it is shown in the appendix \ref{appanqh}, the harmonic approximation
implies that
the eigenstates $|e^{(\mathrm{int})}_n\rg$ depend in fact on two
positive integers $n_\pm$ (we use the notation
$|e^{(\mathrm{int})}_{n_\pm}\rg$). It results
\begin{align}
\lg e^{(\mathrm{int})}_{n_\pm} | \hat{\mathrm{D}} |
e^{(\mathrm{int})}_{n_\pm} \rg &= 0, \\
\nonumber  \lg e^{(\mathrm{int})}_{n_\pm} | \hat{\mathrm{D}}^2 |
e^{(\mathrm{int})}_{n_\pm} \rg &= \frac{1}{2}
(n_+^2+n_-^2+n_++n_-+3) \,.
\end{align}
Therefore, the expectation value $\lg \Psi(t) | \hat{\mathsf{H}} |
\Psi(t) \rg$ of the Hamiltonian $\hat{\mathsf{H}}$, as defined in
\eqref{quhamgentransform},  for $| \Psi(t) \rg =U(\hat{q})( |
\lambda q(t) ,p(t) \rg \otimes | e^{(\mathrm{int})}_{n_\pm} \rg) $
corresponding to the case \ref{BHap} (Born-Huang-like
approximation), reads
\begin{align}
& \lg \Psi(t) | \hat{\mathsf{H}} | \Psi(t) \rg = \\
\notag & \lg \lambda q(t),
p(t) | \left[ \frac{9}{4} \left( \hat{p}^2 + 
\frac{\hbar^2(\frak{K}_1+\chi(n_+, n_-))}{\hat{q}^2} \right) +\right.\\
\notag & \left. 36 \mfn^2
\frak{K}_3 \hat{q}^{2/3}-
E_N(\hat{q}) \right] | \lambda q(t), p(t) \rg  \,,
\end{align}
with
\begin{align}
N = \,&n_+ + n_-,\quad  \textrm{and} \\
\nonumber &\chi(n_+, n_-) =
\frac{2}{9} (n_+^2+n_-^2+n_++n_-+3) \,.
\end{align}
Hence, up to the modification $\frak{K}_1 \mapsto \frak{K}_1 +
\chi(n_+,n_-)$, we recover the previous expression of
Eq.\,\eqref{bornoppenhamred} for the Born-Oppenheimer-like case
\ref{BOap}. From Eq.\;\eqref{bornoppensemi} we deduce the
final expression of the semi-classical Hamiltonian (in the
harmonic approximation of $V_{\mfn}(\beta)$)
\begin{align}
\label{bornhuangsemi}
\check{\mathsf{H}}^{\mathrm{red}}_{n_\pm}(q,p) & = \\
\nonumber & \frac{9}{4} \left( p^2
+ \frac{\hbar^2(\frak{K}_4+ \frak{K}'_4 \chi(n_+,n_-))}{q^2}
\right) +\\
\nonumber &36 \mfn^2
\frak{K}_5 q^{2/3} - \frac{24 \hbar}{q^{2/3}} \frak{K}_6
\mfn (N+1) \,,
\end{align}
where $\frak{K}'_4 = \left( \xi_{10} \right)^2 \,\xi_{20}$.

\section{Semiclassical dynamics} \label{hamcons}

The quantum-corrected constraint  $\check{\mathsf{H}}^{\mathrm{red}}_N(q,p) =0$
may be interpreted as a semiclassical version of the Friedmann equation.
The anisotropic degrees of freedom, averaged at the
quantum level, give rise to the isotropic radiation energy density.
This energy gravitates like common matter and fuels the contraction. It leads
to a supplementary term in the Friedmann equation, shown below.

\subsection{Effective Friedmann-like equation}

Rewritten in terms of the scale factor $a$, the constraint
$\check{\mathsf{H}}_{av}(q,p)=0$ reads
\begin{equation}
\label{newFLRW} \frac{\dot{a}^2}{a^2} + k \frac{c^2}{a^2}+
c^2\ell^{-2} \frac{\frak{K}_4}{a^6} = \frac{8 \pi G}{3 c^2}
\rho(a) \, ,
\end{equation}
where, using the Planck area $\frak{a}_P = 2 \pi G \hbar c^{-3}$,
\begin{align}
k = &\frac{\frak{K}_5
\mfn^2}{4}, \quad \ell =\frac{\mathcal{V}_0}{\frak{a}_P},\\
 \nonumber & \rho(a)
= \mfn \mathcal{V}_0^{-1} \frak{K}_6
\frac{\hbar c (N+1)}{a^4}
\,.
\end{align}
The classical constraint  \eqref{classconstraint} is recovered for
 $\hbar \to 0$. The main features of this quantum corrected model are:
\begin{itemize}
       \item[(i)] The value of the isotropic curvature, $k c^2 a^{-2}$, present in
       closed FRW models, is dressed by the quantization with a constant
       $\frak{K}_5$, which is close enough to $1$ to be ignored in qualitative
       considerations.
       \item[(ii)] The {\it repulsive} potential term proportional to $a^{-6}$, absent
       in classical FRW/BIX models, is generated by the affine CS quantization.
      \item[(iii)]  The energy of the anisotropic oscillations is turned at the quantum
      level into the radiation energy,
       $\rho(a)$. The expression for $\rho(a)$ in terms of the quantum number
       $N$ becomes a poor approximation for high values of $N$, due to
       the breakdown of harmonic approximation. Nevertheless, the dimensional
       analysis shows that the dependence $\rho(a) \propto\mfn \mathcal{V}_0^{-1}\hbar \, c
       \, a^{-4}$ is  correct for $a \to 0$.
\end{itemize}

\subsection{Comparing classical with semiclassical constraint}

The classical Friedmann equation generalized to Bianchi IX geometry reads
\begin{equation}\label{cc1}
\frac{\dot{a}^2}{a^2} + \frac{1}{6} ~^3R-\frac{1}{6}\Sigma^2=0  \,,
\end{equation}
where $~^3R$ and $\Sigma^2$ are respectively the curvature and shear of the spatial sections.
The Bianchi IX curvature \eqref{b9curvature} may be splitted into  isotropic and anisotropic 
parts and  the shear may be expressed in Misner's canonical variables as follows:
\begin{equation}\label{cc2}
~^3R=3\frac{1-V(\beta)}{2a^2},~~~~\Sigma^2=\frac{p_+^2+p_-^2}{24a^6} \, ,
\end{equation}
where the anisotropic part of curvature potential $V(\beta)$, defined by  \eqref{b9potential} 
and \eqref{ani-potential}, reads explicitly as:
\begin{align}\label{h2}
&V(\beta) =  \frac{e^{4 \beta_+}}{3} \times \\
\nonumber  & \left( \left( e^{-6\beta_+} - 2 \cosh (2 \sqrt{3} \beta_-) \right)^2 - 4 \right)+1.
\end{align}
This way we arrive at the following form of \eqref{cc1}:
\begin{equation}\label{cc3}
\frac{\dot{a}^2}{a^2}+\frac{1}{4 a^2}=\frac{1}{6}\,\frac{p_+^2+p_-^2}{24a^6} +\frac{V(\beta)}{4 a^2} \, .
\end{equation}
The left-hand side of \eqref{cc3} contains only the isotropic variables, whereas the right-hand side 
contains all the anisotropic terms. The first and the second term of \eqref{newFLRW}, namely
$\dot{a}^2 \, a^{-2}$ and   $k c^2 \,a^{-2}$, correspond to the first and the second term of \eqref{cc3}, i.e.  to $\dot{a}^2 \, a^{-2}$ and $(2a)^{-2}$, respectively. The third term in \eqref{newFLRW}, 
$c^2\ell^{-2} {\frak{K}_4}{a^{-6}}$, has no corresponding term in the classical equation. It is a purely 
quantum effect induced by the affine CS quantization. The right-hand side of \eqref{newFLRW} describes 
the quantized anisotropic energy, which depends on the scale factor as $ a^{-4}$. It corresponds to the 
energy of anisotropy oscillations in the r.h.s. of \eqref{cc3}, which depends on the scale factor roughly 
as $ a^{-6}$. This difference between the right-hand sides of \eqref{newFLRW} and \eqref{cc3} in 
dependence on $a$ is another purely quantum effect.

Concerning the curvature-like term $\frak{K}_5
\mfn^2 c^2 (2a)^{-2}$ in the semiclassical Eq. \eqref{newFLRW}, we note that it reproduces the classical re-collapse in the evolution of the 
vacuum Bianchi IX model, which results from $S^3$ topology of this model \cite{wald}.
The re-collapse is included in the plots of Fig.\;(\ref{figure3}). The fact
that our semiclassical procedure reproduces this particular classical feature can be viewed 
as the confirmation of the consistency of our approach.

Furthermore, from the semi-classical Lagrangian approach developed in Subsection \ref{appsclag}, we note that any effective matter term $F(q) = \rho(q^{2/3})$ with $\rho(a) =K a^{-n}$ may be incorporated in the Lagrangian of Eq. \eqref{quhamgen}, leading after the action minimization of Eq. \eqref{vmH0} to a supplementary matter term $\tilde{\rho}(a)= \xi K a^{-n}= \xi \rho(a)$, ($\xi >0$), in the r.h.s. of Eq. \eqref{newFLRW}. Therefore adding an effective radiative matter term like $\rho_0 a^{-4}$ (where $\rho_0$ is a positive 
constant) into the r.h.s. of \eqref{newFLRW} 
would bring only a quantitative change in the semiclassical dynamics as it scales exactly as 
quantized anisotropy.

\subsection{Singularity resolution}

Equation \eqref{newFLRW} implies
\begin{equation}\label{conda0}
k c^2 + c^2\ell^{-2}\frac{\frak{K}_4}{a^4}-
\frac{8 \pi G}{3 c^2} a^2 \rho(a) \le 0 \, ,
\end{equation}
which defines the allowed values assumed by $a$. The inequality (\ref{conda0})
can be satisfied only for
\begin{equation}
(N+1)^2 \ge \frac{9}{16} \frac{\frak{K}_4
\frak{K}_5}{\frak{K}_6^2} \,,
\end{equation}
or $(N+1)^2 \ge f(\nu)$ with
\begin{align}
\label{condN}
f(\nu):= &\frac{9}{32} \,
\left(\xi_{02} \right)^4 \left(\xi_{10} \right)^{1/3} \xi_{\frac{2}{3}1} \times\\
\nonumber & \left(\frac{\nu^2}{16} + \frac{\nu}{4} \xi_{21}+ \frac{3 \nu}{8} \xi_{10} + \xi_{20} \right)  \,.
\end{align}
The function $f(\nu)$ is strictly increasing, from $0$ to
$+\infty$, with $f(\nu) \simeq  \frac{9}{512}\nu^2\;$ in the limit $\;\nu \to
\infty$. Therefore for each value of $N$, there exists a bounded
interval $]0,\nu_m(N)]$ for which the condition \eqref{condN}
holds true, with $f(\nu_m(N))=(N+1)^2$.

For large values of $N$, we have $\nu_m(N) \simeq \,
\frac{16 \sqrt{2}}{3} (N+1)$, and $\nu_m(N) \ge \nu_m(0) \simeq 7.4$.\\
Thus, for $\nu \in ]0,\nu_m(0)[$ and for all $N \ge 0$, the
condition \eqref{conda0} is satisfied. We find that $a \in
[a_-,a_+]$, where
\begin{equation}
a_{\pm}^2 =\frac{8 \frak{a}_P}{3\mfn \mathcal{V}_0}  \frac{\frak{K}_6}{\frak{K}_5}
(N+1) \left(1\pm \sqrt{1-\frac{f(\nu)}{(N+1)^2}} \right)
\end{equation}
Therefore, the semiclassical trajectories turn out to be {\it
periodic}, and $a$ is {\it bounded} from below by $a_-$. This
demonstrates that the system does not have the {\it singularity}
that occurs at the classical level of the FRW/BIX model. We note that $a_{\pm}$
is dimensionless as $\mfn \mathcal{V}_0 \propto \mfn^{-2}$ is homogeneous to
an area. Moreover, the volume of the universe, $a^3\mathcal{V}_0$, is independent
of $\mfn$. This proves that our result is physical.

\begin{figure}
\includegraphics[width=6.5cm]{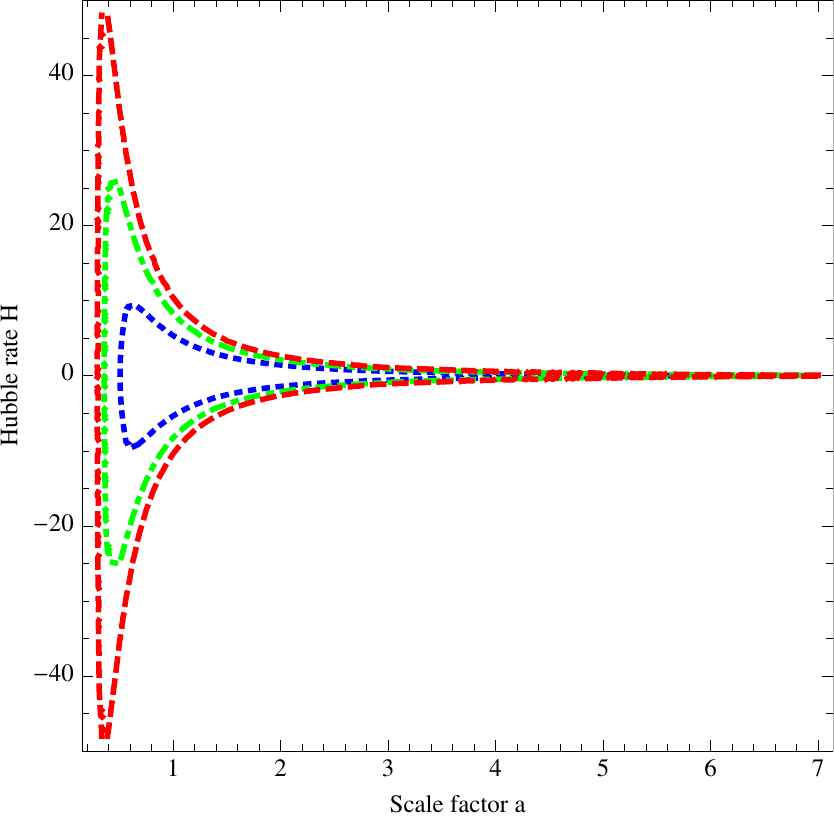}
\caption{Three periodic semiclassical trajectories in the half-plane
$(a,H)$ from Eq.\,\eqref{newFLRW}. They are smooth plane curves.
We use standard units $\frak{a}_P=c=\hbar=1$ and choose $\nu=3$, $\mathcal{V}_0=1$
(so $\mfn=(16 \pi^2)^{1/3}$). Blue dotted curve for $N=0$,
green dotdashed for $N=1$ and red dashed for
$N=2$.  Each plot includes the quantum bounce (at small $a$) and classical recollapse 
(at large $a$). }\label{figure3}
\end{figure}

In Fig.\;\ref{figure3}
we plot a few trajectories in the half-plane $(a,H)$. The
classical closed FRW model is recovered at $\hbar = 0$ and large
values of $\nu$.

{\it Some remarks}:
\begin{itemize}
     \item[(i)] The product $a_- a_+$ is only dependent on $\nu$,
     \begin{equation}
     a_- a_+ = \frac{2 \frak{a}_P}{\mfn \mathcal{V}_0} \sqrt{\frac{\frak{K}_4}{\frak{K}_5}}\,.
     \end{equation}
     \item[(ii)] Specifying $N$ and $\nu=\nu_m(N)$, the relation $a_-=a_+$
     holds true so we have an unusual feature of a stationary universe
     with finite radius.
     \item[(iii)] For $N=0$ and $\nu \in ]0,\nu_m(0)[$, the model
     shows the effect of the `quantum zero point energy' of the anisotropy
     degrees of freedom.
     \item[(iv)] For $\nu \in ]0,\nu_m(0)[$ and all $N
     \ge 0$, the oscillation  period $T$ of the universe is
     \begin{equation}
      T = \frac{4}{\mfn c \sqrt{ \frak{K}_5}}\, a_- \,E\left(1-
      \left(\frac{a_+}{a_-} \right)^2 \right)\,,
\end{equation}
where $E$ is the complete elliptic
integral of the second kind \cite{magnus66}.
\end{itemize}

\section{Beyond adiabatic approximation}\label{consistency}

As we have seen in Section (\ref{bianchiA}), when the isotropy of
the closed FRW universe is perturbed, the universe acquires the
Bianchi IX geometry and a small perturbation inevitably develops
into the chaotic regime, first described in Ref. \cite{BKL1}. To see
whether such behavior is possible at the quantum level, we now go
beyond the adiabatic approximation and allow the number of
particles $N$ to grow as the universe contracts, bounces and
re-expands. We assume that the scale factor is a $c$-number, which
evolves according to the semiclassical constraint \eqref{newFLRW}
for $N=0$. The anisotropy degrees of freedom are quantized as
before but with the possibility to be excited by the
time-dependent background.

We resort to the well-known fact about the harmonic  oscillator
that its classical and quantum dynamics of the basic variables are in one-to-one correspondence and we will work in the Heisenberg picture by
solving classical equations of motion. First, we  find the
semiclassical evolution of $a$ in suitable time parameter, $\tau$.
Then  we use this evolution to define our wave in the
time-dependent background in terms of an oscillator with
``time-dependent mass". Finally, we perform numerical
computations.

\subsection{Evolution of scale factor}
The semiclassical constraint  \eqref{newFLRW} may be written as:
\begin{equation}\label{cons}
H^2 + \frac{\Lambda_1}{a^6}+\frac{\Lambda_0}{a^2} =
\frac{
\Lambda_2}{ a^4}  \, ,
\end{equation}
In what follows we set $\Lambda_0=0$. This simplification removes
the classical re-collapse, the universe expands forever now. This
is a good approximation as we are not interested in the classical
phase but in the quantum one during which the isotropic intrinsic
curvature is assumed negligible. Moreover, we may work in natural
units in which $\Lambda_1\simeq\Lambda_2$ as both terms origin
from quantum theory.

We find from (\ref{cons})
\begin{equation}\label{sol1}
\ud t=\frac{1}{\sqrt{\Lambda_2}}\frac{a^2\ud a}{\sqrt{a^2-\frac{\Lambda_1}{\Lambda_2}}}
\end{equation}
where $t$ is the {\it cosmological time}.

The dynamics of anisotropy  is given by background-dependent Hamiltonian:
\begin{equation}
\mathsf{H}_{\pm}=\frac{1}{2}p_{\pm}^2+144\mfn^2a^4(\tau)\beta_{\pm}^2
\end{equation}
which is a part of the Hamiltonian constraint \eqref{Hamilphys2}-\eqref{Hamilphys2pm}
if the lapse is set to $\mathcal{N}=-12a^3$. The idea of the subsequent
calculations is to treat $a(\tau)$ as a fixed function of time. To obtain
the correct solution $a(\tau)$, we need to adjust (\ref{sol1}) for the choice
of lapse $\mathcal{N}$. We find
\begin{align}
\ud\tau=&\frac{\ud t}{\mathcal{N}}=-\frac{1}{12 \sqrt{\Lambda_2}}\frac{\ud a}{a\sqrt{a^2-
\frac{\Lambda_1}{\Lambda_2}}}\\
\nonumber &=\frac{1}{12 \sqrt{\Lambda_2}}~
\ud\Big[\arcsin\left(\sqrt{\frac{\Lambda_1}{\Lambda_2}}a^{-1}\right)\Big]
\end{align}
hence
\begin{equation}
a(\tau)=\frac{\sqrt{\frac{\Lambda_1}{\Lambda_2}}}{\sin\left(12\sqrt{\Lambda_2}\tau+\frac{\pi}{2}\right)}
\end{equation}
where $\tau\in\left(-\frac{\pi}{24\sqrt{\Lambda_2}},\frac{\pi}{24\sqrt{\Lambda_2}}\right)$
and $a\in\left(+\infty,\sqrt{\frac{\Lambda_1}{\Lambda_2}}\right)$.

\subsection{Excitation of quantum oscillator}

The Hamiltonian under study is
\begin{equation}
\mathsf{H}_{\pm}=\frac{1}{2}p_{\pm}^2+\frac{1}{2}\omega^2(\tau)\beta_{\pm}^2
\end{equation}
where $\omega=12\sqrt{2}\mfn a^2(\tau)$. In what follows we drop $\pm$ for brevity.
The equation of motion reads:
\begin{equation}\label{eq1}
\frac{\ud^2\beta}{\ud\tau^2}=-\omega^2(\tau)\beta
\end{equation}
We will work in the Heisenberg picture, and assume that
\begin{align}
&\hat{\beta}(\tau)=\frac{1}{\sqrt{2}}\left(av^*(\tau)+a^{\dagger}v(\tau)\right),\\
\nonumber &\hat{p}(\tau)=\frac{1}{\sqrt{2}}\left(a\acute{v}^*(\tau)+a^{\dagger}\acute{v}(\tau)\right)
\end{align}
where $a$ and $a^{\dagger}$ are fixed operators and where $v(\tau)$ solves the equation
(\ref{eq1}), i.e.,
\begin{equation}\label{eq2}
\frac{\ud^2v}{\ud\tau^2}=-\omega^2(\tau)v
\end{equation}
We demand the canonical commutation relation and we obtain
\begin{align}
&iI=[\hat{\beta},\hat{p}]=[\frac{1}{\sqrt{2}}\left(av^*(\tau)+a^{\dagger}v(\tau)\right),\\
\nonumber &\frac{1}{\sqrt{2}}\left(a\acute{v}^*(\tau)+a^{\dagger}\acute{v}(\tau)\right)]=
[a,a^{\dagger}]\frac{v^*\acute{v}-v\acute{v}^*}{2}
\end{align}
We find from e.o.m. (\ref{eq2})
\begin{equation}
\frac{\ud}{\ud\tau}\left(v^*\acute{v}-v\acute{v}^*\right)=0
\end{equation}
and we fix $v^*\acute{v}-v\acute{v}^*=2i$. So, $a$ and $a^{\dagger}$ are annihilation
and creation time-independent operators. All time dependence lies in $v(\tau)$.

The Hamiltonian now reads
\begin{align}
& \hat{\mathsf{H}}:=\frac{1}{2}\hat{p}^2+\frac{1}{2}\omega^2(\tau)\hat{\beta}^2\\
\nonumber &=\frac{a^2}{4}
\left((\acute{v}^*)^2+\omega^2(v^*)^2\right)+\frac{(a^{\dagger})^2}{4}\left(\acute{v}^2+
\omega^2v^2\right)+\\
\nonumber & \frac{2a^{\dagger}a+1}{4}\left(|\acute{v}|^2+\omega^2|v|^2\right)
\end{align}
We set the vacuum state $|0\rangle$ for $a,a^{\dagger}$ to minimize the energy at some
initial moment $\tau_0$. It follows that
\begin{equation}\label{ini}
v(\tau_0)=\frac{1}{\sqrt{\omega(\tau_0)}},~~\acute{v}(\tau_0)=i{\sqrt{\omega(\tau_0)}}
\end{equation}
and hence
\begin{equation}
\hat{\mathsf{H}}(\tau_0):=\left(a^{\dagger}a+\frac{1}{2}\right)\omega(\tau_0)
\end{equation}
where $\omega(\tau_0)=12\sqrt{2}\mfn a^2(\tau_0)$.

Now, if we assume $|0\rangle$ to be the initial quantum state, and take into account
the both modes `$\pm$', then the number of particles N at some later time $\tau_1$
can be found with the formula:
\begin{align}
& \langle 0|N(\tau_1)+1|0\rangle=\frac{\langle 0|\hat{\mathsf{H}}_+(\tau_1)+\hat{\mathsf{H}}_-
(\tau_1)|0\rangle}{\omega(\tau_1)}\\
\nonumber & =\frac{|\acute{v}|^2(\tau_1)+
288\mfn^2 a^4(\tau_1)|v|^2(\tau_1)}{24\sqrt{2}\mfn a^2(\tau_1)}
\end{align}
What remains to be solved is (\ref{eq2}) with initial data (\ref{ini}).

\subsection{Numerical results}
\begin{figure}
\includegraphics[width=0.45\textwidth]{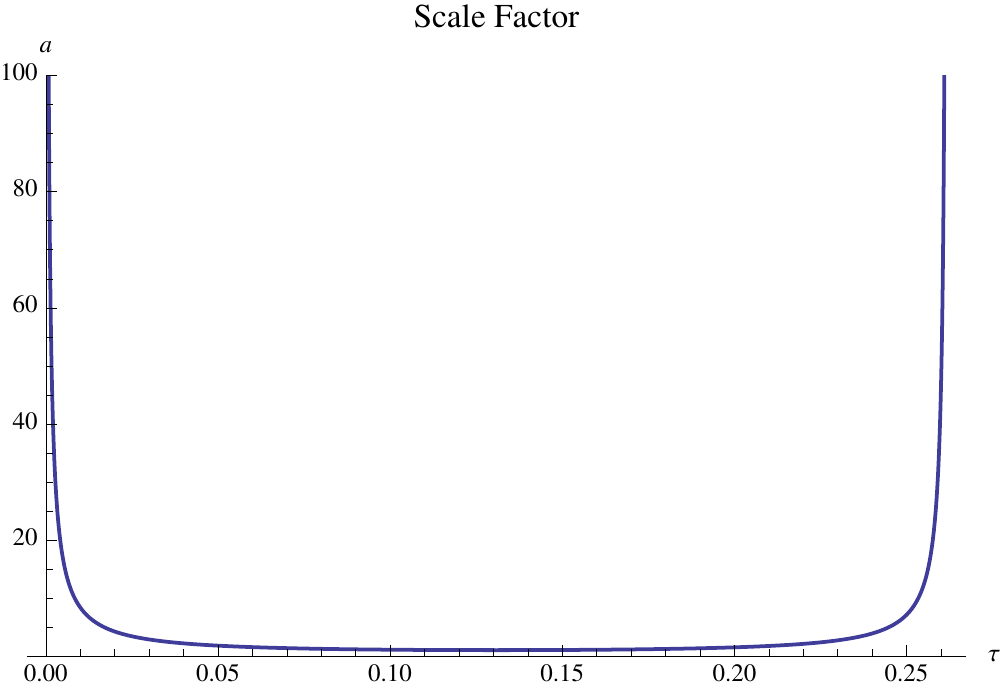}
\includegraphics[width=0.45\textwidth]{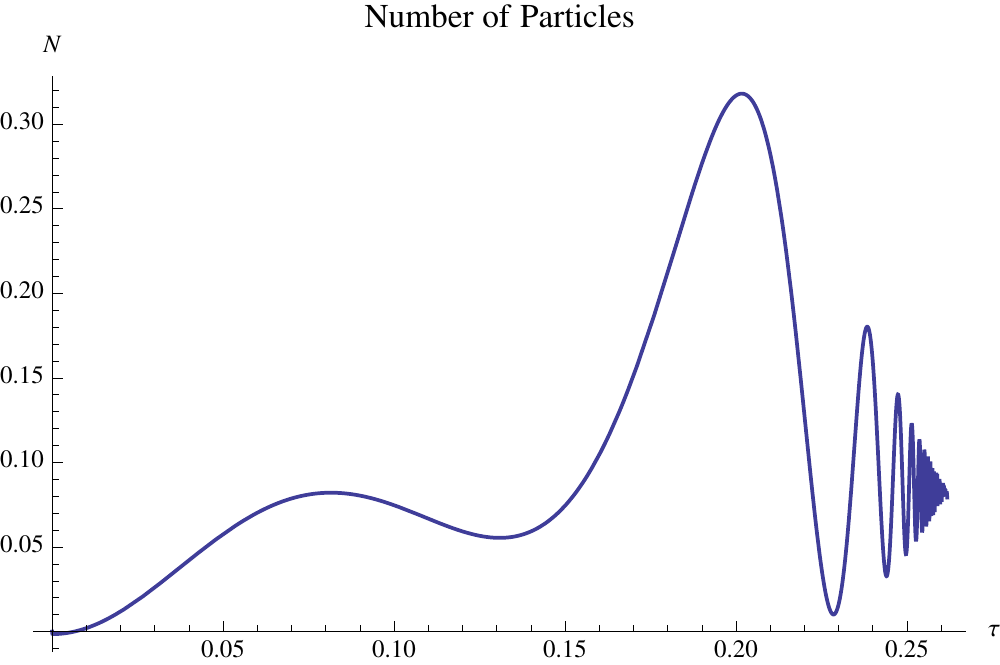}
\caption{The evolution of the number of particles $N$ and the
scale factor $a$ from the contracting phase through the
semiclassical bounce to the expanding phase. We fix
$\Lambda_1=1=\Lambda_2$, $\mfn=1$, and
$a(\tau_0)=10^{4}$.}\label{figure4}
\end{figure}

We fix units so that $\Lambda_1=1=\Lambda_2$ and also $\mfn=1$. So
the universe contracts from $a=\infty$, bounces at $a=1$ and
re-expands to $a=\infty$. We set the initial data as
$a(\tau_0)=10^{4}$. In Fig.\;(\ref{figure4}) we plot the number
of particles and the scale factor versus time. We also find that
varying initial $a$ from $10^{4}$ to $10$ does not affect the
number of particles produced, which never exceeds $0.35$.

\section{Conclusions}

\label{conclusion}

In the present article we have examined the  quantum dynamics of the vacuum
Bianchi IX model, the Mixmaster universe. The
Hamiltonian constraint, which consists of isotropic and
anisotropic variables, has been quantized. This split of variables is crucial
both for
implementing our procedure and interpreting the result.
Suitable coherent states, namely the ACS, have been employed to obtain some
insight into the involved
quantum dynamics of isotropic background. Making use of adiabatic
approximation, we have  identified the eigenstates for the oscillating
anisotropy at its lowest excitation levels. Our procedure,
developed by qualitative arguments and based on reasonable and tractable approximations, i.e. the replacement of the anisotropy potential with the harmonic potential and the Born-Oppenheimer-type evolution, is  validated by considerations outside the adiabatic approximation.

The main features of our quantum model are the following: (i) the
singularity avoidance due to a repulsive term regularizing the
singular spacetime geometry; (ii) the reduced contraction rate of
the universe due to suppressed growth of the energy of anisotropy
at the quantum level as it becomes frozen in a fixed quantum state;
(iii) the stability of quantum Friedmann-like state in the quantum
Bianchi IX model both in the contraction and expansion phase. We
emphasize that the anisotropic oscillations have a non-zero
ground energy level, which means that there is no true quantum FRW
state.

The resolution of the singularity is due to the repulsive potential generated by
the ACS quantization as in \cite{Bergeron:2013ika}. We note that our approach
to `quantizing singularity' is very natural. It was developed within an appealing
probabilistic interpretation of quantization procedures and, in a sense, it extends
what is usually meant by canonical quantization (see Appendix \ref{affquant}).
Our approach is universal in the sense that it removes the singularity from
anisotropic models, recently shown to be true also for Bianchi I models in \cite{BDGM2}.

In our framework, due to the Born-Oppenheimer approximation, the
anisotropy degrees of freedom are assumed to be in a  quantum
eigenstate. This property leads to the radiation energy density.
In the near future, we will focus on developing our scheme to include
transitions between different energy levels by more detailed computations.
The Born-Huang-like
approximation is just a refined version of the adiabatic approximation as
shown in Subsection (\ref{appsclag}). Qualitatively,
the Born-Huang-like approximation does not change the behavior of
the semi-classical Hamiltonian obtained within the
Born-Oppenheimer-like approximation, because the former generates
a supplementary positive term $\propto q^{-2}$ that only
renormalizes the term already present (due to our affine CS
quantization). On the other hand,  if we apply only canonical quantization to the
system, then no term in $q^{-2}$ is present at the beginning in
the Born-Oppenheimer-like approximation. In that case the
Born-Huang-like approximation generates a new term in the
semi-classical behavior. Since this repulsive term is responsible
of the resolution of the singularity, we can say that with or without
ACS quantization, the resolution of the singularity holds true in the
case of the Bianchi IX model within the framework of Born-Huang
approximation.

We emphasize that our approach is completely different from
the one of Misner's, which was based on the so called steep wall
approximation, discussed in Appendix (\ref{appanqh}). In the steep
wall approximation, there is no tendency for quantum probability
to be peaked in the minimum of potential, and therefore there
would probably be no quantum suppression of anisotropy as it
occurs for the real potential revealed in the harmonic
approximation of our paper.

We have shown that the  wave remains in its lowest energy states
during the quantum phase. Even  beyond adiabatic approximation
there is no significant excitation of the wave's energy level.
It is interpreted that the quantum  FRW universe, unlike its
classical version, is dynamically stable with respect to small
isotropy perturbation. Thus, supplementing the FRW Hamiltonian
with the zero-point energy originating from quantization of
anisotropic degrees of freedom provides a quantum version of
the Friedmann model which could be used for a study of the
earliest Universe.

\acknowledgments
P.M. was supported by MNiSW Fellowship ``Mobilno\'s\'c Plus".

\appendix

\section{Affine coherent state quantization}

\label{affquant}

Coherent state quantization is a particular approach pertaining to
what is named in \cite{bergaz14} integral quantization. When a
group action is involved in the construction, one can insist on
covariance aspects of the method. A detailed presentation of the
subject is given in \cite{bergaz14} and in Chapt.\;11 of
\cite{G2000}. In this appendix, we give a short compendium of this
approach before particularizing to the integral quantization
issued from the affine group representation.

\subsection{Covariant integral quantizations}
\label{covintquant}

Lie group representations \cite{barracz77} offers a wide range of
possibilities for implementing integral quantization(s).  Let $G$
be a Lie group with left Haar measure $\mathrm{d}\mu(g)$, and let
$g\mapsto U\left(g\right)$ be a unitary irreducible representation
(UIR) of $G$ in a Hilbert space $\mathcal{H}$. Consider a bounded
operator $\mathsf{M}$ on $\mathcal{H}$ and suppose that the
operator
\begin{equation}
\label{boundR}
R:=\int_{G}\mathsf{M}\left(g\right)\mathrm{d}\mu\left(g\right),\
\mathsf{M}\left(g\right):=U\left(g\right)\mathsf{M}U^{\dagger}\left(g\right)\,,
\end{equation}
is defined in a weak sense. From the left invariance of $\mathrm{d}\mu(g)$
we have
\begin{equation}
\label{comRU}
U\left(g_{0}\right)RU^{\dagger}\left(g_{0}\right)=\int_{G}\mathsf{M}\left(g_{0}g\right)
\mathrm{d}\mu\left(g\right)=R\,,
\end{equation}
so $R$ commutes with all operators $U(g)$, $g\in G$. Thus, from
Schur's Lemma, $R=c_{M}I$ with
\begin{equation}
\label{cM}
c_{M}=\int_{G}tr\left(\rho_{0}\mathsf{M}\left(g\right)\right)\mathrm{d}\mu\left(g\right)\,,
\end{equation}
where the unit trace positive operator $\rho_{0}$ is chosen in
order to make the integral converge. This family of operators
provides the resolution of the identity on $\mathcal{H}$.
\begin{equation}
\int_{G}\mathsf{M}\left(g\right)\mathrm{d}\nu\left(g\right)=I,\qquad \mathrm{d}\nu\left(g\right)
:=\frac{\mathrm{d}\mu\left(g\right)}{c_{M}}\,.\label{eq:resolution}
\end{equation}
and the subsequent quantization of complex-valued functions (or distributions, if well-defined) on $G$
\begin{equation}
\label{quantgr}
f\mapsto A_{f}=\int_{G}\, \mathsf{M}(g)\, f(g)\,\mathrm{d}\nu(g)\,,
\end{equation}
This linear map, function $\mapsto$ operator in $\mathcal{H}$, is
covariant in the sense that
\begin{equation}
\label{covarG}
U(g)A_{f}U^{\dagger}(g)=A_{\mathfrak{U}(g)f}\,.
\end{equation}
In the case when $f\in L^{2}(G,\mathrm{d}\mu(g))$, the quantity
$(\mathfrak{U}(g)f)(g^{\prime}) :=f(g^{-1}g^{\prime})$ is the regular
representation.

A semi-classical analysis of the operator $A_f$ can be implemented
through the study of  lower symbols. Suppose that $\mathsf{M}$ is
a density, i.e. non-negative unit-trace, operator $\mathsf{M}=
\rho$ on $\mathcal{H}$. Then the operators $\rho(g)$ are also
density, and this allows to build a new function $\check{f}(g)$ as
 \begin{equation}
\label{lowsymb}
\check{f}(g)\equiv \check{A_f}:=\int_{G}\, tr(\rho(g)\,\rho(g^{\prime}))\,
f(g^{\prime})\mathrm{d}\nu(g^{\prime})\,.
\end{equation}
The map $f\mapsto \check{f}$ is a generalization of the Berezin or
heat kernel transform on $G$ (see \cite{hall06} and references
therein). One observes that the value $\check{f}(g)$ is the average about $g$ of the original 
$f$ with respect to the probability distribution on $(G, \mathrm{d}\nu(g^{\prime}))$ defined by $g^{\prime} \mapsto tr(\rho(g)\,\rho(g^{\prime}))$.  

Let us consider the above procedure in the case of square
integrable UIR's and rank one $\rho$. For a square-integrable UIR
$U$ for which $\left\vert \psi\right\rangle $ is an admissible
unit vector, i.e.,
\begin{equation}
\label{cpsi}
c(\psi):=\int_{G}\mathrm{d}\mu(g)\,|\left\langle \psi\right\vert U\left(g\right)\left\vert
\psi\right\rangle |^{2}<\infty\,,
\end{equation}
the resolution of the identity is obeyed by the coherent states
$\left\vert \psi_{g}\right\rangle =U(g)\left\vert
\psi\right\rangle$, in a generalized sense, for the group $G$:
\begin{align}
\label{residsq}
\int_{G}\rho(g)  \mathrm{d}\nu\left(g\right)=I\ ,\quad & \mathrm{d}\nu\left(g\right)=
\frac{\mathrm{d}\mu\left(g\right)}{c(\psi)}\, , \\
\nonumber  \rho(g)&=\left\vert
\psi_g\right\rangle \left\langle \psi_g\right\vert \,.
\end{align}

\subsection{The case of the affine group} \label{affG}

As the complex plane is viewed as the phase space for the motion of
a particle on the line, the half-plane is viewed as the phase
space for the motion of a particle on the half-line. Canonical quantization of the plane 
is covariant  in the sense that it respects the translation symmetry of the plane through its 
Weyl-Heisenberg group extension. Our approach to the quantization of the half-plane follows the same principle of covariance with respect the affine  group structure of this geometry.

Let  the upper half-plane $\Pi_{+}:=\{(q,p)\,|\, p\in\mathbb{R}\,,\, q>0\}$
be equipped with the  measure $\mathrm{d}q\mathrm{d}p$.
Together with the multiplication
\begin{equation}
\label{multaff}
(q,p)(q_{0},p_{0})=(qq_{0},p_{0}/q+p),\, q\in\mathbb{R}_{+}^{\ast},\, p\in\mathbb{R}\, ,
\end{equation}
the unity $(1,0)$ and the inverse
\begin{equation}
\label{invaff}
(q,p)^{-1}= \left(\frac{1}{q}, -qp \right)\, ,
\end{equation}
$\Pi_{+}$ is indeed viewed as the affine group Aff$_{+}(\mathbb{R})$ of
the real line, and the  measure $\mathrm{d}q\mathrm{d}p$ is left-invariant with
respect to this action. The affine group Aff$_{+}(\mathbb{R})$ has two non-equivalent
UIR \cite{gelnai47,aslaklauder68}.
Both are square integrable and this is the rationale behind \textit{continuous
wavelet analysis} (see references in \cite{G2000}). The UIR $U_{+}\equiv U$
is realized in the Hilbert space $\mathcal{H}=L^{2}(\mathbb{R}_{+}^{\ast},\mathrm{d}x)$:
\begin{equation}
U(q,p)\psi(x)=(e^{ipx}/\sqrt{q})\psi(x/q)\,.\label{affrep+}
\end{equation}
By adopting the integral quantization scheme described above, we
restrict  to the specific case of rank-one density operator or
projector $\rho=|\psi\rangle\left\langle \psi\right|$ where $\psi$
is a unit-norm state in
$L^{2}(\mathbb{R}_{+}^{*},\mathrm{d}x)$ which should be also in 
$L^{2}(\mathbb{R}_{+}^{*},\mathrm{d}x/x)$ (also called
``fiducial vector\textquotedblright{} or
``wavelet\textquotedblright{}). The action of UIR $U$ produces all
affine coherent states, i.e. wavelets, defined as $|q,p\rangle\
=U(q,p)|\psi\rangle$.

Due to the irreducibility and square-integrability of the UIR $U$,
the corresponding quantization reads as
\begin{equation}
\label{quantfaff}
f\ \mapsto\ A_{f}=\int_{\Pi_{+}}f(q,p)|q,p\rangle\langle q,p|\dfrac{\mathrm{d}q\mathrm{d}p}{2\pi c_{-1}}\,,
\end{equation}
which arises from the resolution of the identity
\begin{equation}
\label{affresunit}
\int_{\Pi_{+}}|q,p\rangle\langle q,p|\,\dfrac{\mathrm{d}q\mathrm{d}p}{2\pi c_{-1}}=I\,,
\end{equation}
where
\begin{equation}
\label{cgamma}
c_{\gamma}:=\int_{0}^{\infty}|\psi(x)|^{2}\,\frac{\mathrm{d}x}{x^{2+\gamma}}\,.
\end{equation}
Thus, a necessary condition  to have \eqref{affresunit} true is
that $c_{-1} < \infty$, which explains  the needed square integrability of $\psi$ w.r.t. $\ud x/x$ implies $\psi(0) = 0$, a well-known requirement in wavelet analysis.

The map \eqref{quantfaff} is covariant with respect to the unitary
affine action $U$:
\begin{equation}
\label{covaff}
U(q_0,p_0) A_f U^{\dag}(q_0,p_0) = A_{\mathfrak{U}(q_0,p_0)f}\, ,
\end{equation}
with
\begin{align}
\label{covaff}
 \left(\mathfrak{U}(q_0,p_0)f\right)(q,p) &=
f\left((q_0,p_0)^{-1}(q,p)\right)\\
\nonumber &= f\left(\frac{q}{q_0},q_0(p -p_0) \right)\, ,
\end{align}
$\mathfrak{U}$ being the left regular representation of the affine
group. In particular, this (fundamental) property is used to prove
Eq.\;\eqref{lam}.

To simplify, we pick a real fiducial vector. For the simplest functions,  the affine CS quantization produces the following operators
\begin{align}
\label{quantqp}
A_{p}= -i\frac{\partial}{\partial x}\equiv \hat{p}\,,\quad & A_{q^{\beta}}=\frac{c_{\beta-1}}{c_{-1}}
\,\hat{q}^{\beta}\,,\\
\nonumber  \hat{q} f(x) &=xf(x)\,.
\end{align}
Whereas $A_q=(c_0/c_{-1}) \hat{q}$ is self-adjoint, the operator $\hat{p}=A_p$ is symmetric but has no self-adjoint extension.
We check that this affine  quantization is, up to a multiplicative constant, canonical,
$[A_q,A_p]= i c_0/c_{-1} I $.\\
We obtain the exact canonical rule (i.e. $A_q=\hat{q}$ and $A_p=\hat{p}$) by imposing $c_0=c_{-1}$. This  simply corresponds to a rescaling of the fiducial vector $\psi$ as $\psi_1(x) = \psi(x/\mu)/ \sqrt{\mu}$ with $\mu = c_0/c_{-1}$.  In the remainder we assume that this rescaling has been done and therefore $c_0(\psi)=c_{-1}(\psi)$.

The quantization of the product $qp$ yields:
\begin{equation}
\label{quantqpdil}
A_{qp}= \frac{c_0}{c_{-1}}\frac{\hat{q} \hat{p} + \hat{p}\hat{q}}{2} = \frac{\hat{q} \hat{p} + \hat{p}\hat{q}}{2} \equiv \, D\,,
\end{equation}
where $D$ is the dilation generator.  As one of the two generators
(with $\hat{q}$)  of the UIR $U$ of the affine group, it is essentially
self-adjoint.

The quantization of the kinetic energy gives
\begin{align}
\label{qkinener}
A_{p^{2}}&=\hat{p}^{2}+K\hat{q}^{-2}\,,\\
\nonumber  K&=K(\psi)=\int_{0}^{\infty}
(\psi^{\prime}(u))^{2}\,u\frac{\mathrm{d}u}{c_{-1}}.
\end{align}
Therefore, wavelet quantization prevents a quantum free particle
moving on the positive line from reaching the origin. It is well
known that the operator $\hat{p}^{2}=-\ud^{2}/\ud x^{2}$ in
$L^{2}(R_{+}^{\ast},\mathrm{d}x)$ is not essentially self-adjoint,
whereas the above regularized operator, defined on the domain of
smooth function of compact support, is essentially self-adjoint
for $K\geq3/4$ \cite{GezKiR85}. Then, quantum dynamics of the free motion is unique.

As usual, the semi-classical aspects are included in the phase
space. The quantum states and their dynamics have phase space
representations through wavelet symbols. For the state
$|\phi\rangle$ one has
\begin{equation}
\label{Phisym}
\Phi(q,p)=\langle q,p|\phi\rangle/\sqrt{2\pi}\,,
\end{equation}
with the associated probability distribution on phase space given
by
\begin{equation}
\label{rhophi}
\rho_{\phi}(q,p)=\dfrac{1}{2\pi c_{-1}}|\langle q,p|\phi\rangle|^{2}.
\end{equation}
Having the (energy) eigenstates of some quantum Hamiltonian $H$ at
our disposal, we can compute the time evolution
\begin{equation}
\label{rhophiev}
\rho_{\phi}(q,p,t):=\dfrac{1}{2\pi c_{-1}}|\langle q,p|e^{-iH t}|\phi\rangle|^{2}
\end{equation}
for any state $\phi$.
The map \eqref{lowsymb} yielding lower symbols from classical $f$ reads in
the present case (supposing that Fubini's theorem holds):
\begin{align}
\label{afflowsymb}
\nonumber &\check{f}(q,p)= \frac{1}{\sqrt{2\pi}c_{-1}} \int_0^{\infty}
\frac{\ud q^{\prime}}{qq^{\prime}}\,\int_0^{\infty}\ud x \,\int_0^{\infty}\ud
x^{\prime}  \\
& e^{ip(x^{\prime}-x)}
F_p(q^{\prime},x-x^{\prime})\,\times\\
\nonumber & \psi\left(\frac{x}{q}\right)\,
\psi\left(\frac{x}{q^{\prime}}\right)\,\psi\left(\frac{x^{\prime}}{q}\right)\,
\psi\left(\frac{x^{\prime}}{q^{\prime}}\right)\,,
\end{align}
where $F_p$ stands for the partial inverse Fourier transform
\begin{equation}
\label{parcoure}
F_p(q,x)= \frac{1}{\sqrt{2\pi}}\int_{-\infty}^{+\infty} e^{ipx} f(q,p)\, .
\end{equation}
For functions $f$ depending on $q$ only, expression \eqref{afflowsymb} simplifies
to a lower symbol  depending on $q$ only:
\begin{align}
\label{lowfq}
\check{f}(q)= &\frac{1}{c_{-1}}\int_0^{\infty}\frac{\ud q^{\prime}}{qq^{\prime}}\,
f(q^{\prime}) \int_0^{\infty}\ud x\,\\
\nonumber & \times \psi^2\left(\frac{x}{q}\right)\,\psi^2\left(\frac{x}{q^{\prime}}\right)\,.
\end{align}
For instance, any power of $q$ is transformed into the same power up to a constant factor
\begin{equation}
\label{powq}
q^{\beta} \mapsto  \check{q^{\beta}}= \frac{c_{\beta-1}c_{-\beta-2}}{c_{-1}} \, q^{\beta}\, .
\end{equation}
Note that $c_{-2} = 1$ from the normalization of $\psi$. \\

We notice that $\check{q} = c_0 c_{-3} (c_{-1})^{-1} q$. If we choose the fiducial vector
such that $c_0=c_{-1}$ in order to obtain the canonical rule, it remains $\check{q} =c_{-3} q$.
Using a $q$-rescaling of the coherent states in the definition of the symbols $\check{A}_f$
like $\check{A}_f = \lg \lambda q, p |\, A_f \, | \lambda q,p \rg$ allows to
obtain $\check{q}=q$ if we choose $\lambda = 1/c_{-3}$.
\\
Other important symbols are:
\begin{equation}
\label{symbp}
p \mapsto \check{p}= p\, ,
\end{equation}
\begin{align}
\label{symbp2}
p^2 \mapsto \check{p^2} &= p^2 + \frac{\mathrm{c(\psi)}}{q^2}\, ,\\
\nonumber  \mathrm{c(\psi)}
& = \int_0^{\infty}\left(\psi^{\prime}(x)\right)^2\,\left(1+\frac{c_0}{c_{-1}} x \right)\,\ud x\,.
\end{align}
\begin{equation}
\label{symbp2}
qp \mapsto \check{qp}= \frac{c_0 c_{-3}}{c_{-1}}qp\,
\end{equation}
Another interesting formula in the  semi-classical context
concerns the Fubini-Study metric derived from the symbol of total
differential $d$ with respect to parameters $q$ and $p$ affine
coherent states,
\begin{equation}
\label{dercs}
\lg q,p|d|q,p\rg= iq\, dp\int_0^{\infty}(\psi(x))^2\, x\, \ud x = iq\, dp \,c_{-3} \,.
\end{equation}
and from norm squared of $d|q,p\rg$,
\begin{align}
\label{norm2}
\Vert d|q,p\rg\Vert^2 &= c_{-4}\,q^2\, dp^2 + L\, \frac{dq^2}{q^2}\, ,\\
\nonumber  L &=
\int_0^{\infty}\ud x\, x^2\, (\psi^{\prime}(x)^{\prime})^2 - \frac{1}{4}\,.
\end{align}
With Klauder's notations \cite{klauderscm}
\begin{align}
\label{fubstud}
d\sigma^2(q,p):=& 2\left\lbrack \Vert d |q,p\rg\Vert^2 - \vert \lg q,p|d|q,p\rg\vert^2\right\rbrack \\
\nonumber 
= & \, 2\left((c_{-4}-c_{-3}^2)  q^2\, dp^2 + L\, \frac{dq^2}{q^2}\right)\,.
\end{align}

\section{Quantum anisotropies}
\label{appanqh}

\subsection{General setting}

The Hamiltonian $\hat{\mathsf{H}}_q$ of Eq.
\eqref{eq:hamilpm} reads
\begin{equation}
\label{hamqB}
\hat{\mathsf{H}}_q= \frak{K}_2\frac{\hat{p}_+^2+\hat{p}_-^2}{q^2}+
36 \frak{K}_3 q^{2/3}V_{\mfn}(\beta)\,,
\end{equation}
where
\begin{align}
&V_{\mfn}(\beta) = \frac{\mfn^2}{3} \times\\
\nonumber & e^{4\beta_+}\left(\left[2\cosh(2\sqrt{3}\beta_-)-
e^{- 6\beta_+}\right]^2-4\right) +\mfn^2\,.
\end{align}
More explicitly, we have for $\hat{\mathsf{H}}_q$ the
expression:
\begin{equation}
\hat{\mathsf{H}}_q = \frac{2\frak{K}_2 \hbar^2}{q^2} \, \hat{\mathcal{E}}(q) \,,
\end{equation}
with
\begin{align}
\label{hamqB1}
\hat{\mathcal{E}}(q) = - \frac{1}{2} \Delta + \chi(q) &\,
V_{\mfn}(\beta), \quad 
\Delta=\partial_{\beta_+}^2+\partial_{\beta_-}^2, \\
\nonumber \chi(q) &= \frac{18
\frak{K}_3}{\frak{K}_2 \hbar^2}
q^{8/3} \,.
\end{align}
$V_{\mfn}(\beta)$ possesses an absolute minimum for $\beta_\pm=0$,
and near this minimum we have
\begin{equation}
\label{eq:approxupot} V_{\mfn}(\beta) = 8 (\beta_+^2+\beta_-^2) +
o(\beta_\pm^2) \,.
\end{equation}
As mentioned above, $V_{\mfn}(\beta)$ and therefore
$\hat{\mathcal{E}}(q)$ possess the symmetry ${\sf C}_{3v}$. This
group has three irreducible representations usually called $A_1$,
$A_2$ and $E$. Therefore the eigenstates of $\hat{\mathcal{E}}(q)$
can be classified according to these representations.

\subsection{Harmonic approximation}

\label{section:harmonic}

Using  \eqref{eq:approxupot} we obtain
\begin{equation}
\hat{\mathcal{E}}(q) \simeq - \frac{1}{2} \Delta + 8 \chi(q) \,(\beta_+^2+\beta_-^2).
\end{equation}
Introducing the quantum numbers $n_\pm=0,1,\dots$, corresponding
to the independent harmonic Hamiltonians in $\beta_+$ and
$\beta_-$, we deduce the harmonic approximation of the eigenvalues $e(n_+,n_-)$ of $\hat{\mathcal{E}}(q)$:
\begin{equation}
e(n_+,n_-) \simeq 4 \sqrt{ \chi(q)}(n_++n_-+1),
\end{equation}
which gives the approximation of the eigenvalues $E_N(q)$
of $\hat{\mathsf{H}}_q$, with $N = n_++n_-$,
\begin{equation}
E_N(q) \simeq \frac{24 \hbar}{q^{2/3}} \mfn \sqrt{2 \frak{K}_2 \frak{K}_3}
\, (N+1) \,.
\end{equation}

\subsection{Steep wall approximation}

As mentioned above, taking into account the ${\sf C}_{3v}$
symmetry and the exponential walls of the potential, we can
approximate $V_{\mfn}(\beta)$ by an equilateral triangular box as
shown on the figure \ref{figure3}. The interest of this
approximation is that it preserves the symmetry ${\sf C}_{3v}$ of
the potential and it \red{also} possesses an explicit solution in terms of
eigenstates and eigenvalues \cite{WaiLi1985}, \cite{WaiLi1987},
\cite{Gaddah2013}.

The size of the triangle is a free parameter that must be somehow
adjusted, e.g., through some variational method.

\begin{figure}[!ht]
\includegraphics[scale=0.8]{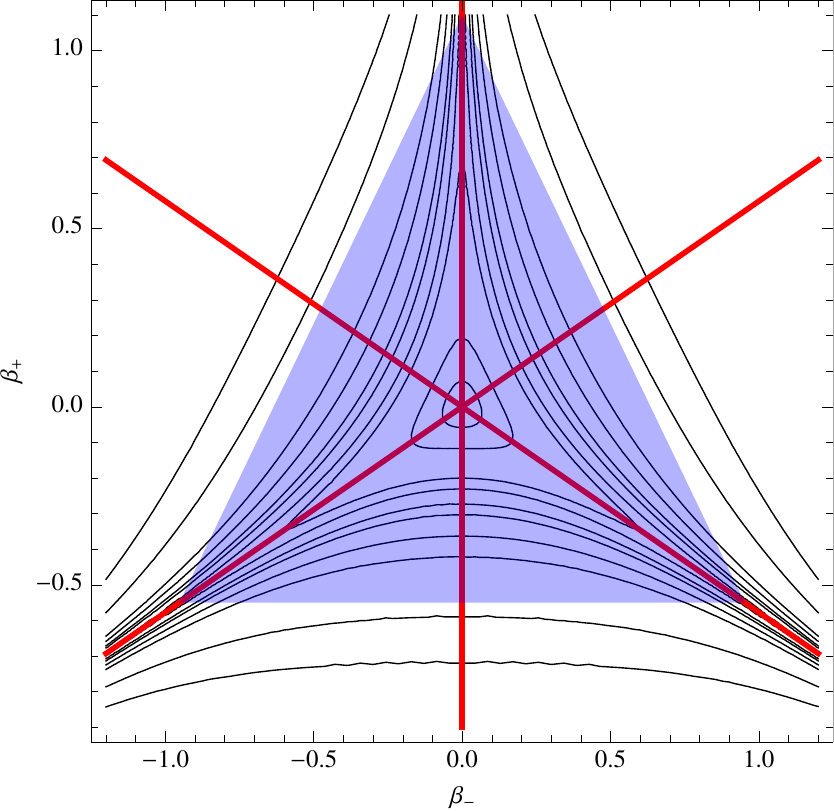}
\caption{Contour plot of $V(\beta)$ near its minimum. Red
lines: the three ${\sf C}_{3v}$ symmetry axis $\beta_-=0$,
$\beta_+=\beta_-/\sqrt{3}$,  $\beta_+=-\beta_-/\sqrt{3}$. A
possible triangular box approximation in blue.} \label{figure5}
\end{figure}

Let us denote by $b$ the side length of the equilateral triangle
box $T_b$ of Fig.\;\ref{figure5}, and let us denote by $U_T$ the
potential equal to $0$ inside the triangle and equal to
$+\infty$ outside. The stationary Schr\"odinger equation
$-\frac{1}{2} \Delta \psi =e^{(T)} \psi$ with the
Dirichlet boundary conditions has explicit solution
\cite{WaiLi1985, WaiLi1987, Gaddah2013}. The eigenvalues
$e^{(T)}_{m,n}$ with $m=0,1,2,\dots$ and $n=1,2,\dots$ are given
by\footnote{We have changed the parametrization of
\cite{WaiLi1985} to have independent integers.}
\begin{equation}
e^{(T)}_{m,n} = \frac{8 \pi^2}{3 b^2} \left( \frac{m^2}{3} +
n^2+m n \right) \,.
\end{equation}

The ground state energy is $e^{(T)}_{0,1} $. The corresponding
normalized ground state wave function\footnote{We have modified
the solution $\psi_{0,1}$ of \cite{WaiLi1985} in order to take
into account the different origin and the orientation of the
triangle.} $\psi_{0,1}$ reads \cite{WaiLi1985}
\begin{align}
\psi_{0,1}(\beta) &= \sqrt{\frac{8}{3 \sqrt{3} b^2}} \left(
\sin \left( \frac{4 \pi \beta_+}{b \sqrt{3}} + \frac{2 \pi}{3}
\right) + \right.\\
\nonumber & \left. 2 \sin \left( \frac{2 \pi \beta_+}{b \sqrt{3}} + \frac{
\pi}{3} \right) \cos \frac{2 \pi \beta_-}{b} \right),
\end{align}
where $\beta_\pm \in T_b $.

Assuming that the harmonic approximation
(Sec. \ref{section:harmonic}) yields a fairly good approximation  to the ground
energy of $\hat{\mathcal{E}}(q)$, the length parameter $b$ can be  estimated by imposing that 
the two ground states coincide up to some small correction
\begin{equation}
e^{(T)}_{0,1} \approx 4 \sqrt{\chi(q)} \,.
\end{equation}
This relation yields
\begin{equation}
b \approx \sqrt{\frac{2 \pi^2}{3 \sqrt{\chi(q)}}} \,.
\end{equation}
This leads to a new approximation $e(m,n)$ of the eigenvalues of
$\hat{\mathcal{E}}(q)$ as
\begin{equation}
e(m,n) \simeq  4 \sqrt{\chi(q)} \left( \frac{m^2}{3}
+ n^2+m n \right) \,.
\end{equation}
The eigenvalues $E_N$ of $\hat{\mathsf{H}}_q$ are still
given formally by
\begin{equation}
E_N(q) \simeq \frac{24 \hbar}{q^{2/3}} \mfn \sqrt{2 \frak{K}_2 \frak{K}_3} \,
(N+1) \,,
\end{equation}
but now $N$ does not reduce to a simple integer. It  is given by
\begin{align}
N &=  \frac{m^2}{3} + n^2+m n - 1, \quad \text{with} \\
\nonumber & m=0,1;\dots, \, \, n=1,2,\dots
\end{align}
The last remark concerns the actual discrete spectrum of the Hamiltonian \eqref{hamqB} (or \eqref{hamqB1}). Clearly, it interpolates between the spectrum of the isotropic 2d-harmonic oscillator, which is linear in $N$ and the spectrum  of the triangular steep wall which behaves like squared integers. This is reminiscent of the 1d-Schr\"{o}dinger operator with a symmetric P\"{o}schl-Teller potential for which the two types, harmonic and infinite square well,  of approximations hold (see for instance \cite{PT01}). In this article we have privileged the ``harmonic'' side of the anisotropy potential for obvious reasons, but  it seems to us that a rigorous complete  mathematical analysis of the spectral properties of the operator \eqref{hamqB} is still lacking.

\end{document}